\documentclass{amsart}
\usepackage{graphicx}
\usepackage{graphicx}
\usepackage{amssymb}
\usepackage{amsfonts}
\usepackage{amsmath}
\usepackage{amsthm}
\newtheorem{Theorem}{Theorem}[part]
\newtheorem{Definition}{Definition}[part]

\newtheorem{Assumption}{Assumption}[part]
\newtheorem{Lemma}{Lemma}[part]

\newtheorem{Remark}{Remark}[part]

\newcommand{\be}{\begin{equation}}
\newcommand{\ee}{\end{equation}}

\numberwithin{equation}{section}

\begin{document}
\title{Additive Habit Formation: Consumption in Incomplete Markets with Random Endowments}

 \author{Roman Muraviev\\
 Department of Mathematics and RiskLab\\
 ETH Zurich\\
 }%

\address{
 Department of Mathematics and RiskLab, ETH, Zurich 8092, Switzerland.
 \newline
 {e.mail: roman.muraviev@math.ethz.ch}}

 \date{July 23, 2011.}
\begin{abstract}
We provide a detailed characterization of the optimal consumption stream
for the additive habit-forming utility maximization problem, in
a framework of general discrete-time incomplete markets and random endowments.
This characterization allows us to derive the monotonicity and concavity 
of the optimal consumption as a function of wealth, for several important 
classes of incomplete markets and preferences. These results yield 
a deeper understanding of the fine structure of the optimal consumption and 
provide a further theoretical support for the classical conjectures of Keynes \cite{KE}. 
\end{abstract}

\maketitle \markboth{R. Muraviev}{Optimal Consumption with Additive
Habits}
\renewcommand{\theequation}{\arabic{section}.\arabic{equation}}
\pagenumbering{arabic}

\section{Introduction}
\label{intro}
The class of habit-forming expected utility models have become increasingly popular in financial 
economics in recent years. 
While time-separable expected utility models are routinely
found to be inconsistent with experimental evidence on choice under uncertainty
(see Plott \cite{PLOT}), habit preferences tend to be efficient in explaining certain empirical phenomena,
as for instance, the celebrated equity premium puzzle (see for example Abel \cite{A}, Campbell and Cochrane \cite{CH}, 
and Constantinides \cite{CO}). A further beneficial feature of the habit preferences is based
on remarkably appealing grounds that are intuitively reasonable from both an economic and psychological
viewpoint. Individuals who consume portions of their wealth over time are expected
to develop habits which will have a decisive impact on their subsequent consumption behavior.
In particular, the relative desire to consume may be increased if one has become accustomed to
high levels of consumption.
The paradigm of habit-forming utility functions captures the above observations by
incorporating the impact of past consumption patterns on the individuals' current and
future policy resolutions. A broad range of works are devoted to the study of habit-formation
optimization problems in various contexts and applications (Chan and Kogan \cite{CK}, Chapman \cite{CHAPMAN}, 
Detemple and Karatzas \cite{DK}, Detemple and Zapatero \cite{DZ1,DZ2}, Eglezos and Karatzas \cite{EK},
and Heaton \cite{HEA}).

In the current paper, we provide a detailed characterization of the optimal consumption stream
for the additive habit-forming utility maximization problem, in
a framework of general discrete-time incomplete markets and random endowments.
This characterization allows us to derive the monotonicity and concavity 
of the optimal consumption as a function of wealth, for several important 
classes of incomplete markets and preferences.

Utility maximization problems in incomplete markets with the presence of random endowments are known to be particularly
challenging to handle in both discrete and continuous time, even for standard, time-separable preferences. Since the pioneering 
work of Merton \cite{MERTON}, 
this problem has attracted the attention of many authors (see, e.g., He and Pearson \cite{HP1,HP2}, 
Karatzas, Lehoczky et. al. \cite{KLSX}, R\'{a}sonyi and Stettner \cite{RS}, and Malamud and Trubowitz \cite{MT}). 
However, even though existence and uniqueness of the optimal consumption has been established in a very general setting 
(see e.g., Karatzas and Zitkovic \cite{KZ},
and the references therein), very little is known about the precise structure of the optimal consumption. 
As emphasized in the survey paper of Zariphopoulou \cite{Z}, further efforts are
to be made in order to capture the qualitative structure of the optimal consumption policy. 

The presence of habits makes the problem more complicated due to the natural time-inseparability of the underlying preferences. 
In the present paper, we study an individual's life-cycle consumption problem in a fairly general discrete-time setting.
A decision maker is represented by a stream of (unspanned) random endowments and an additive habit-forming
utility function. Agents are imposed to consume in an addictive fashion: Since the utility function is defined on the positive half-line, 
consumption can never fall below the benchmark level. The specification of habits in our
model involves a combination of an internal-external additive mechanism. Namely, the index indicating the standard of living
of an individual is a weighted average of individual's past consumption plus some exogenous (stochastic) factor. The agent
is aiming to maximize the preference functional by trading in an arbitrary incomplete financial market.
In the above framework, we provide a novel characterization of the optimal consumption stream by exploiting the concept of
an aggregate state price density, introduced by Malamud and Trubowitz \cite{MT}. We analyze some
important special cases and study particular qualitative properties of the consumption stream.

To the best of our knowledge, this paper is the first one to study the habit-forming utility maximization problem in the general
framework of incomplete markets and random endowments. Furthermore, unlike the traditional
dynamic programming approach, our methodology allows us to explicitly characterize the optimal consumption stream
for such a general setting. The approach introduced in this work can be employed in a future research dealing with habits in
discrete-time models. Finally, our results allow us to address several important economic questions such as 
the monotonicity and concavity of the optimal consumption. In the standard case of time-separable preferences, these properties have been investigated 
by   
Carroll and Kimball \cite{CK0} and Malamud and Trubowitz \cite{MT}, providing a theoretical foundation for the conjecture of Keynes \cite{KE} 
that the marginal propensity to consume is diminishing. Our new techniques allows us to extend these results to the case of habit-forming preferences. 
This extension is by no means trivial as the nature of first order conditions is significantly more complicated due to the time-inseparability of the 
habit-forming preferences. 

We now outline the contents of this paper. Section 2 deals with preliminaries.
In subsection 2.1 we introduce the model of an incomplete financial market. The aggregate state price density and other
related concepts are introduced in subsection 2.2.
In subsection 2.3, we formulate the habit-forming utility maximization problem and provide a first-phase
solution (Theorem \ref{existence}) in terms of the aggregate state price density.
Section 3 is concerned with a delicate analysis of the utility maximization problem in various
well-known models of financial markets and habit-preferences. This section can serve as an
introduction to the more complex ideas appearing later in Sections 4 and 5.
Subsection 3.1 provides a solution to the optimization problem with preferences represented by a habit-forming power (constant relative risk aversion) 
utility, arbitrary incomplete markets and deterministic endowments. In subsection 3.2, we consider a habit-forming exponential
utility maximization problem in the case where agents can only trade one period riskless bonds. Finally, in subsection 3.3 we solve the utility 
maximization problem
in complete markets and arbitrary habit-forming preferences.
Section 4 is devoted to the study of monotonicity, and to the establishment of an extended characterization of the consumption for a large class
of incomplete markets. First, we explain in subsection 4.1
why standard dynamic programming methods cannot be directly applied to this 
monotonicity problem. Then, in subsection 4.2, the main result of Section 4 (Theorem \ref{main1}) is presented:
The monotonicity feature and an explicit recursive scheme determining the optimal consumption is established
for arbitrary incomplete markets with a
deterministic interest rate, and for idiosyncratically incomplete markets (see Definition \ref{def1}),
in the setting of general additive habit-forming preferences.
In Section 5, we study the concavity of the optimal consumption as a function of wealth. 
That is, we show that the richer an individual is, the smaller is the portion of the wealth consumed by him, approving the hypothesis of 
Keynes \cite{KE}.
Subsections 5.1 and 5.2 are devoted to some surprising
counter-examples illustrating that concavity can only be anticipated for time-consistent power utility functions.
Finally, in subsection 5.3, we formulate the main result of Section 5 (Theorem \ref{main3}), showing the concavity property
for idiosyncratically incomplete markets, and incomplete markets of type $\mathcal{C}$ with a deterministic interest rate
(see Definition \ref{marketc}).

\section{Setting and Preliminary Results}

The uncertainty in our model is captured by a finite probability space $(\Omega, \mathcal{G}, P
) $ and a filtration $ \mathcal{G}_0 := \{ \phi, \Omega \} \subseteq \mathcal{G}_1 \subseteq ... \subseteq \mathcal{G}_T := \mathcal{G}, $
where each sigma-field $\mathcal{G}_k $ corresponds to the information revealed up to the period $k$.
In the current paper, adaptedness of stochastic processes is always meant with respect to
$\mathcal{G} := \left( \mathcal{G}_k \right)_{ k = 0 , ... , T}, $ unless otherwise is stated.
For each $k$, $ L^{2} \left( \mathcal{G}_k \right) $ denotes the (finite-dimensional) space of
 all $\mathcal{G}_k$-measurable random variables endowed with the inner product
$ \langle X,Y \rangle := E \big[X \cdot Y \big], $ for $X, Y \in L^{2}( \mathcal{G}_k )$.
Each sigma-algebra $\mathcal{G}_k $ is generated by a certain partition of $\Omega$, that is,
\[
\mathcal{G}_k = \sigma \big( B^{ (k) }_{1},..., B^{ (k) }_{N_k}
\big),
\]
where $B^{ (k) }_{1}, ... ,B^{ (k) }_{ N_k }$ are disjoint subsets
of $\Omega$, $ \bigcup_{i=1}^{N_k} B^{ (k) }_{i} = \Omega$ and $
N_k = \dim \left( L^{2} \left( \mathcal{G}_k \right) \right) $.
Furthermore, each partition is finer than the previous one, i.e.,
$ B^{ (k) }_{j} = \bigcup_{i \in I^{(k)}_j } B^{ (k + 1 ) }_{i},$
for $k=0 , ... , T-1,$ where $ I^{(k)}_j \subseteq \left\{ 1 , ...
, N_{k+1} \right\}$ are disjoint subsets, for $ j = 1 , ... ,
N_k$. Note that a random variable $X$ is
$\mathcal{G}_k$ measurable if and only if $X$ is constant on each
set $ B^{(k)}_{i} $, for $ i = 1,..., N_k$. For each $ X \in
L^{2}(\mathcal{G}_k)$, we consider the associated
\textbf{multiplication operator} $ T_{X}: L^{2} \left(
\mathcal{G}_T \right) \to L^{2} \left( \mathcal{G}_T \right) $
defined by $ T_{X}(Y) =  X  Y.$ In the standard basis of $
L^{2} \left( \mathcal{G}_T  \right) $, the operator $T_{X}$ is a
diagonal matrix with entries equal to $ X|_{B^{ ( k ) }_{1}} , ...
, X|_{B^{ ( k ) }_{ N_k }} $ that appear according to a certain
multiplicity. We will \textbf{not} distinguish between $X$ and
$T_{X}$ in the forthcoming sections. Given a function
$ f (X, Y) : \mathcal{O} \times V \to L^{2}(\mathcal{G}_T) ,$
where $\mathcal{O} \subseteq L^{2}(\mathcal{G}_k)$ is an open set (in the
$L^{2}(\mathcal{G}_k)$-topology), for $k=0,...,T$, and $V$ is some non-empty
subset of $ L^{2} ( \mathcal{G}_0 ) \times ... \times L^{2} ( \mathcal{G}_T ) $,
we denote by
\[
\frac{ \partial f }{ \partial X } \big|_{ (X_0 ,Z_0 ) } Y_0 = \lim_{ \varepsilon \to 0}
\frac{ f( X_0 + \varepsilon Y_0 , Z_0 ) - f( X_0 , Z_0 ) }{ \varepsilon } ,
\] 
the directional derivative (G\^ateaux differential) of $f$ at the point $ (X_0 , Z_0 ) \in  \mathcal{O} \times V,$ applied on the
vector $Y_0 \in L^{2}( \mathcal{G}_k ) .$ In many cases, the operator $ \frac{ \partial f }{ \partial X } $
will be a multiplication operator, and thus $ \frac{ \partial f }{ \partial X } $ will be treated as a random variable.
Derivatives of higher order are defined in a similar manner.

\subsection{The Financial Market}

We adopt a standard discrete time model in the spirit of Chapter
2 in Duffie \cite{Duffie}. The financial market
consists of $N$ risky securities and one riskless bond. There are
$T+1$ periods: $0,...,T$. At each date $ k = 0, ... ,  T-1$,
each security $i =1 , ... ,  N$ is available for trading (selling or buying) at the price $S^{i}_{k}$ and pays
a dividend $d^{i}_{k+1}$ in the next period $k+1.$ A riskless bond
bought in the period $ k = 0 , ... ,  T-1 ,$ pays an interest rate
$r_{k+1} $ in the next period $k+1$. The \textbf{price process} is
an $N+1$ dimensional positive adapted process given by $ S_k = ( 1
, S^{1}_k,...,S^{N}_k )$, $ k = 0 , ... , T-1.$ We assume that no
trading is executed in the last period $T$, and incorporate this
assumption by imposing $S_T = (0,...,0) \in {R}^{ N + 1 }$.
The \textbf{dividend processes} is an $N+1$-dimensional adapted
process denoted by $d_k =( r_{k} , d^{1}_k,...,d^{N}_k)$, for $ k=
1,..., T-1,$ and $d_T = ( 1 + r_{T}, d^{1}_T,...,d^{N}_T)$.
The interest rate $ \left( r_k \right)_{ k = 1 , ... , T} $ is assumed
to be a non-negative predictable process. We emphasize that $ \left( d_k \right)_{k=1,...,T-1}$ and $d_T$
differ in the first coordinate due to the assumption that $S_T$ vanishes. Moreover,
the first coordinate in $\left( d_k \right)_{k=1,...,T}$ and
$\left( S_k \right)_{k=0,...,T-1}$ is reserved for the riskless bond.
A \textbf{trading strategy}, or portfolio, is an $N+1$ dimensional adapted process
$ \pi_k = ( \phi_k, \pi^{1}_k , ... , \pi^{N}_k ) ,$ $ k = 0 ,
... , T$. Here, $\phi_k$ and $\pi^{i}_k$ represent the
respective shares of the riskless bond and security $i$, held
during the period of time $[ k , k+1 )$.  We set further
$\pi_{-1}= (0,...,0) \in R^{ N + 1 } $ and $\pi_T = (0,...,0) \in R^{ N + 1 }$. 
One notes that the standard
assumption regarding the predictability of the portfolio is
omitted due to a shift in the index of the price process. For a
given trading strategy $\pi$, the associated \textbf{investment}
process $I^{\pi} := (I^{\pi}_k )_{ k = 0,...,T-1 }$ and the
associated \textbf{financial wealth} processes $W^{\pi} :=
(W^{\pi}_k )_{ k = 1,...,T }$ are defined respectively by
\[
 I^{\pi}_k :=  \phi_k + \sum_{i=1}^{N} \pi^{i}_k
 S^{i}_k = \pi_k \cdot S_k,
\]
and
\[
 W^{\pi}_k := \phi_{k - 1 }  (1 + r_k ) + \sum_{i=1}^{N} \pi^{i}_{ k -1}  (S^{i}_k + d^{i}_k) = \pi_{k-1} \cdot \left( S_k + d_k \right),
\]
where $\cdot$ denotes the standard inner product in ${R}^{N+1}$. 
A \textbf{state price density (SPD)} is
an adapted process $ \left( R_k \right)_{ k = 0,...,T}$ that satisfies
\[
 R_k  S^{i}_{ k } = E \left[ R_{ k + 1 } ( S^{i}_{ k+1 } + d^{i}_{ k+1 } ) | G_k
 \right],
\]
and
\[
 R_k = E \left[ R_{ k + 1 }  ( 1 + r_{ k+1 } ) | G_k \right],
\]
for all $ i = 1 , ... , N $ and $ k = 0 , ... ,  T-1.$
\begin{Remark}
(i) In our setting, a SPD is allowed to take negative values.
\newline
(ii) Given an investment process $(I^{\pi}_k)_{ k=0,...,T-1}$,
a wealth process $(W^{\pi}_k )_{ k=0,...,T}$ and an arbitrary SPD
$(R_k )_{ k=0,...,T}$, the following relation
\begin{equation}
 R_{k}  I^{\pi}_{k}  = E \left[ R_{ k+1 }  W^{\pi}_{ k+1 } | \mathcal{G}_{ k } \right]
\label{inv_wealth}
\end{equation}
holds for all $  k = 0 , ... , T-1.$
\label{remark000}
\end{Remark}

The following is assumed throughout the whole paper.

\begin{Assumption}
There is no arbitrage (NA) in the market. That is, if, $ W^{\pi}_{k} - I^{\pi}_{k} \geq 0 ,$ $P-$a.s for
all $k=0,...,T$ and some portfolio $ \pi $, then, $ W^{\pi}_{k} - I^{\pi}_{k} = 0 $, $P-$a.s
for all $k=0,...,T$.
\end{Assumption}

We recall the standard classification of financial markets.

\begin{Definition} A market is complete if every adapted process 
$(Y_k)_{k=1,...,T}$
is replicable, i.e., there exists a trading strategy $\pi$ such
that $ Y_k = I^{\pi}_k - W^{\pi}_k$ for all $ k = 1 , ... ,  T.$
Otherwise, the market is called incomplete.
\end{Definition}

Finally, we state the following classical result in a discrete time setting.

\begin{Theorem} The NA condition is equivalent to the existence of a positive SPD.
A NA market is complete if and only if there exists a unique (up to a multiplication
by a constant) positive SPD.
\label{NA_completeness}
\end{Theorem}
\textbf{Proof of Theorem \ref{NA_completeness}} See the proof of Theorem 3.3 in Dallang, Morton and Willinger \cite{DMW},
and the proof of Corollary 4.1 in Taqqu and Willinger \cite{TW}. $\qed$

\subsection{The Aggregate State Price Density}

The current subsection is devoted to the introduction of certain important notions
that will be crucial for carrying out our analysis. We start with the following definition.

\begin{Definition} (i) The financial wealth space, or payoff space at period $k$ is given by
$$ \mathcal{L}_k = \{ W^{\pi}_{k} = (S_k + d_k)  \pi_{k-1} | \pi_{k-1} \in L^{2} \left( \mathcal{G}_{k-1} \right)  \} , $$
for all $ k = 1 , ... , T,$ and $\mathcal{L}_0 = \{ 0 \}. $
\newline
(ii) We denote by
$$ P^{k}_{\mathcal{L}} : L^{2}(\mathcal{G}_T) \to
\mathcal{L}_k $$ the orthogonal projection of the space
$L^{2}(\mathcal{G}_T)$ onto the payoff space $\mathcal{L}_k ,$ for
all $ k = 0 , ... ,  T.$
\end{Definition}

\begin{Remark}
(i) Since riskless bonds are available for trading at each period, it follows that
$ L^{2} \left( \mathcal{G}_{k-1} \right)
\subseteq \mathcal{L}_k \subseteq L^{2} \left( \mathcal{G}_k
\right)$, for all $ k = 1 ,... , T$.
\newline
(ii) In contrast to conditional expectations, the orthogonal projection
$P^{k}_{\mathcal{L}}$ is not necessarily positivity preserving (i.e., not mapping non-negative random
variables to non-negative ones). In fact, $P^{k}_{\mathcal{L}}$ is positivity
preserving if and only if $\mathcal{L}_k = L^{2} \left(
\mathcal{H}_k \right) $ for some sigma-algebra $
\mathcal{G}_{k-1} \subseteq \mathcal{H}_k \subseteq \mathcal{G}_k $ (See Theorem 3.2
in Malamud and Trubowitz \cite{MT}). In the latter case, the market is called
\textit{market of type $\mathcal{C}$} (see Sections 4.2 and 5.3 for certain results
involving this type of markets).
\end{Remark}

The following properties will be used extensively in the sequel.

\begin{Lemma}
(i) For every $X,Y \in L^{2}(\mathcal{G}_T)$ and $ k = 1 , ... , T
$, we have
$$ E \left[ P_{\mathcal{L}}^{k} \left[ X \right]  Y \right] =  E \left[ X  P_{\mathcal{L}}^{k} \left[ Y \right] \right] .$$
(ii) For every $k=1,...,T,$ $Y \in L^{2}(\mathcal{G}_T)$ and $X \in
L^{2}(\mathcal{G}_{k-1})$, we have
\[
P^{k}_{\mathcal{L}} \left[ X  Y \right] = X P^{k}_{\mathcal{L}}
\left[ Y \right].
\]
(iii) For all $X , Y \in L^{2}(\mathcal{G}_T)$, $ k= 0 , ... , T-1 $
and $ m > k $, we have
\[
 E \left[ P^{m}_{\mathcal{L}} \left[ X \right]  Y | \mathcal{G}_{k} \right] = E \left[ X  P^{m}_{\mathcal{L}} \left[ Y \right] | \mathcal{G}_{k} \right].
\]
(iv) If $P^{k}_{ \mathcal{L} } \left[ X  V \right] = 0$, for some
positive random variable $X \in L^{2} \left( G_T \right) $ and
some $V \in \mathcal{L}_k,$ then $V=0.$
\label{teclma1}
\end{Lemma}
\textbf{Proof of Lemma \ref{teclma1}.} (i) 
The assertion follows from the fact that $P^{k}_{\mathcal{L}}$ is a self-adjoint operator. \\
(ii) First, note that $ X  P^{k}_{\mathcal{L}} \left[ Y \right]
\in \mathcal{L}_k $ by definition. Therefore, the statement
becomes $P^{k}_{\mathcal{L}} \left[ X  Y - X P^{k}_{\mathcal{L}}
\left[ Y \right] \right] = 0 ,$ or equivalently,
 $ E \left[ \left( X  Y - X
P^{k}_{\mathcal{L}} \left[ Y \right] \right)  Z \right] = 0 ,$ for
all $Z \in \mathcal{L}_k .$ Since $X  Z \in \mathcal{L}_k,$ part
(i) implies that $ E \left[  X  Z  P^{k}_{\mathcal{L}} \left[ Y
\right] \right] = E \left[ X  Y  Z \right].$ This completes the
proof of
(ii). \\
(iii) We need to prove that $ E \left[ P^{m}_{\mathcal{L}} \left[ X
\right] Y Z \right]= E \left[ P_{ \mathcal{L}
}^{m} \left[ Y \right] X Z \right] $, for all $Z \in L^{2}(
\mathcal{G}_k)$. It follows by (ii) that $ Z P^{m}_{\mathcal{L}} \left[ X \right] =
P^{m}_{\mathcal{L}} \left[ Z X \right] $. Therefore, by combining this with (i) we
obtain that $ E \left[ P_{ \mathcal{L} }^{m} \left[ X
\right]  Y  Z \right] = E \left[ P_{ \mathcal{L} }^{m}
\left[   X Z \right]   Y  \right] = E \left[ P_{ \mathcal{L} }^{m}
\left[   Y \right]   X  Z  \right],$ as required. \\
(iv) Note that by (i) we have $ 0 = E \left[ P^{k}_{ \mathcal{L} }
\left[ X  V \right]  V \right] = E \left[ X  V^2 \right]$. Since $
X > 0 $, we conclude that $V=0.$  \qed 
${}$
\newline
The concept of aggregate SPD is a fundamental tool in the present
work.

\begin{Theorem}
There exists a unique SPD $ ( M_k )_{ k = 0,...,T }$ such that $ M_k \in
\mathcal{L}_k,$ for all $ k= 1 , ... ,  T$. Moreover,
\[
M_{k} = \prod_{ l  = 1}^{k} P_{\mathcal{L}}^{ l } \left[ \frac{R_{ l }}{R_{ l - 1 }} \right],
\]
for all $ k = 1 , ... ,  T,$ where $ ( R_k )_{ k= 0 ,...,T} $ is
an arbitrary positive SPD. In particular, $( M_k )_{ k = 0,...,T
}$ is independent of the choice of $( R_k )_{k=0,...,T}$. The
process $( M_k )_{ k = 0,...,T }$ is called the \textit{aggregate
SPD}.
\label{spd}
\end{Theorem}
\textbf{Proof of Theorem \ref{spd}.} See the proof of Lemma 2.5 in Malamud and Trubowitz \cite{MT}. $\qed$
\newline
\newline
Note that the aggregate SPD is not necessarily a positive process.
For simplicity, we will consider only financial markets that
satisfy the following convention.
\begin{Assumption}
The aggregate SPD does not vanish, i.e., $M_k \neq 0,$ $P-$a.s,
for all $ k = 1 , ... ,  T.$
\end{Assumption}

\subsection{Utility Maximization with Additive Habits}
The decision maker in our model is characterized by an endowment
stream $ \epsilon := (\epsilon_k)_{k=0,...,T} $ which is assumed to
be a non-negative adapted process, and a habit-forming utility
function $U$. The individual's objective is to maximize her
utility function from consumption by selecting a suitable
investment policy. Given a trading strategy $ \pi $, the
corresponding consumption process labeled by $ c^{ \pi} := (c^{
\pi}_k)_{k=0,...,T},$ is defined by
\begin{equation}
c_k = c^{\pi}_k := \epsilon_k + W^{\pi}_k - I^{\pi}_k.
\label{consumption}
\end{equation}
The preference functional of the agent is given by
\[
 U( c_0 , ... , c_T ) := \sum_{ k=0}^{T}  E \left[ u_{k} \left( c_k - \sum_{ l = 0}^{k-1} \beta^{(k)}_{l}  c_{ l } - h_k   \right) \right],
\]
where $ \beta := ( \beta^{(k)}_{l} )_{ l = 0 , ... , k-1 }$, for
$k=1,...,T,$ are non-negative constants that represent the strength
of the habit formation affect on the individual, and $ ( h_{k} )_{ k =
0 , ... , T}$ is a non-negative adapted process that might be
interpreted as a quantity which measures habits of exogenous type, or as a
mending factor that appears due to a noisy estimation of the
consumption's past history. For notational convenience, we set $
\beta^{(0)}_{-1} = h_{0} = 0 $ and $\beta^{(k)}_k = 1$, for $ k =
1 ,..., T$. We assume that each function $u_{k} : [ 0, \infty )
\to {R} $ is a strictly increasing, concave, $C^2$-smooth
and satisfies the Inada conditions: $ u'_k(0) := \lim_{x \to 0}
u'_k(x) = \infty ,$ and $ u'_k( \infty) := \lim_{x \to \infty }
u'_k (x) = 0.$ Note that the latter convention implies that
feasible consumption streams are non-negative and that individuals
are addictive, that is, $ c_k \geq \sum_{ l = 0 }^{ k-1 } \beta^{
( k ) }_{ l }  c_{ l } + h_k \geq 0 ,$ for all $ k = 0 , ... ,  T.
$ For each consumption plan $ ( c_k )_{ k= 0 , ... , T}  $ we
define the associated perturbed consumption process $ ( c ( k ,
\beta , h ) )_{ k = 0 , ... , T }$ by
\begin{equation}
c ( k , \beta , h ) := c_k - \sum_{ l = 0}^{k-1} \beta^{(k)}_{l}  c_{ l } - h_k.
\label{per_consumption}
\end{equation}
Note that the utility functions in the current setting are not restricted by the assumption that $u_{k}(x) = e^{-\rho k} u(x),$ 
for all $k=0,...,T$, where $u(x)$ is some function satisfying the Inada conditions. 
This leads to time-inconsistency, as for instance in Ekeland and Lazrak \cite{EL}, and Karp \cite{KARP}. The following subset of 
$L^{2}(\mathcal{G}_0) \times
L^{2}(\mathcal{G}_1) \times ... \times L^{2}(\mathcal{G}_T)$ will
be referred to as the set of budget constraints
\[
B \left( \epsilon, \beta, h \right) :=
\]
\[
 \left\{ (c_0,...,c_T)  |  c_k = c^{ \pi }_k \ \mbox{for some portfolio} \ \pi \ \mbox{and} \
c ( k , \beta , h ) \geq 0 \ \mbox{for all} \ k=0,...,T \right\}.
\]
We assume throughout the paper that the set $ B \left( \epsilon, \beta, h \right)$ has a 
non-empty interior. The utility maximization problem of the agent is
\begin{equation}
 \sup_{ (c_0,...,c_T) \in B \left( \epsilon, \beta, h \right) } U(c_0, ... , c_T).
\label{utility_max}
\end{equation}
The following proposition provides a convenient parametrization
of the set of budget constraints.
\begin{Lemma}
Let $ ( c_k )_{ k = 0,...,T}$ be an adapted process that satisfies the
condition $ c ( k , \beta , h) \geq 0 ,$ for all $ k = 0 , ...
,  T.$ Then, $ ( c_k )_{ k = 0,...,T}$ is a feasible consumption
stream, i.e., $ (c_0,...,c_T) \in B \left( \epsilon, \beta, h
\right)$, if and only if there exists a unique wealth process
$(W_{k})_{k=0,...,T}$, i.e., $W_{k} \in \mathcal{L}_k$ for all $
k=0 , ... , T$ and $W_{T+1}=0,$ such that
\begin{equation}
 c_{k} = \epsilon_{k} + W_{k} - E \left[ \frac{M_{k+1}}{M_{k}}  W_{k+1} | \mathcal{G}_k \right],
\label{budgetc1}
\end{equation}
for all $ k = 0 , ... ,  T$. Furthermore, if $\pi$ is some trading
strategy such that $ c_k = c^{\pi}_k$ for all $k=0,...,T$, then
$W_k = W^{\pi}_{k}$ and $ E \left[ \frac{M_{k+1}}{M_{k}} W_{k+1} |
\mathcal{G}_k \right] = I^{\pi}_k,$ for all $k=0,...,T$. The
wealth process $ ( W_{ k} )_{ k = 0,...,T}$ is uniquely determined
by
\begin{equation}
W_{k} = \sum_{l=k}^{T} E \left[ \frac{M_l}{M_k}  \big( c_l - \epsilon_l
\big) | \mathcal{G}_k \right] \label{budgetc2} ,
\end{equation}
for all $ k = 1 , ... , T.$
\label{bc_parametrization}
\end{Lemma}
${}$
\newline
This lemma is analogous to Lemma 2.7 in Malamud and Trubowitz \cite{MT}. We present its proof for 
the reader's convenience. 
\newline
\newline
\textbf{Proof of Lemma \ref{bc_parametrization}}
The first part of the statement follows by (\ref{inv_wealth}). Let us
now show that (\ref{budgetc2}) holds by backward induction.
The case $k=T$ follows from (\ref{budgetc1}). Assume
that (\ref{budgetc2}) holds for all $k > m $, and let us show
it for $ k = m $. By combining equation (\ref{budgetc1}) with the induction
assumption, we get that $ W_{
m } = c_{ m } - \epsilon_{ m } + E [ \frac{M_{ m
+1}}{M_{ m }}  W_{ m +1} | \mathcal{G}_{ m } ] = c_{
m } - \epsilon_{ m } + \sum_{ l = m + 1 }^{ T } E [ \frac{
M_{ m + 1 } }{ M_{ m } }  E [ \frac{ M_l }{M_{ m + 1 } }
 (c_l - \epsilon_l ) | \mathcal{G}_{ m  +  1 } ] |
\mathcal{G}_{ m } ] = \sum_{ l = m  }^{T} E [ \frac{M_l }{M_{
m }}  (c_l - \epsilon_l ) | \mathcal{G}_m ]. $ This accomplishes the proof.  \qed
${}$
\newline
The next statement establishes a first step in the characterization of the
optimal consumption stream in the setting of general incomplete
markets.
\begin{Theorem}
There exists a unique solution to the utility maximization problem (\ref{utility_max}). The optimal
consumption stream $ ( c^{*}_k )_{k=0,...,T} $ and the corresponding financial wealth process
$ (W^{*}_{k})_{k=0,...,T} $ are uniquely determined as the solutions of the following system of equations:
\begin{equation}
 P^{ k }_{ \mathcal{L} } \left[ \frac{ R_{k}( c_0,...,c_T ) }{ R_{k - 1}( c_0,...,c_T ) } \right] =  \frac{ M_{k} }{ M_{k-1}
 },
\label{first1}
\end{equation}
for $ k = 1 , ... , T $, where
\[
R_{k}( c_0,...,c_T )  :=  u'_{k} \left( c(  k , \beta , h )
\right) - \sum_{ m = k + 1}^{T} \beta^{(m)}_{ k }  E \left[ u'_{m}
\left( c(  m , \beta , h )  \right) \big| \mathcal{G}_{ k }
\right] ,
\]
$ ( c(  k , \beta , h ) )_{ k = 0 , ... , T}  $ is defined in (\ref{per_consumption}),
\begin{equation}
c_{k} = \epsilon_{k} + W_{k} - E \left[ \frac{M_{k+1}}{M_{k}}  W_{k+1} | \mathcal{G}_k \right],
\label{first2}
\end{equation}
and $W_k \in \mathcal{L}_k$, for all $ k = 0,...,T$. The process $
( R_{k}( c^{*}_0,...,c^{*}_T ) )_{ k = 0 , ... , T } $ is a
positive SPD. 
\label{existence}
\end{Theorem}
\textbf{Proof of Theorem \ref{existence}} The compactness of
$B(\epsilon, \beta, H)$ (which follows by the NA assumption) combined with the continuity
and the strict concavity of the function $U(c_0,...,c_T)$ implies existence and uniqueness.
Set $ c^{*} ( k , \beta , h_k ) := c^{*}_k - \sum_{ l = 0}^{k-1} \beta^{(k)}_{l}  c^{*}_{ l } - h_k$
and observe that the Inada condition $u'_k(0)=\infty$ implies that $ c^{*} ( k , \beta , h_k ) > 0,$ $P-$a.s, for all $ k =
0 , ... , T $.
We turn to showing that $(c^{*}_k)_{ k = 0 , ... , T}$ is
determined as the unique solution of the system of equations
(\ref{first1}) and (\ref{first2}). Note that Lemma
\ref{bc_parametrization} implies that equations (\ref{first2})
can replace the budget constraint $ ( c^{*}_0 , ... , c^{*}_T) \in
B \left( \epsilon, \beta, h \right)$. Next, due to the fact that $
(c^{*}_0,...,c^{*}_T) \in int \left( B \left( \epsilon, \beta, h
\right) \right) $ and that a local maximum of a strict concave
function is a global maximum,  it follows that
$(c^{*}_0,...,c^{*}_T)$ is determined as the unique solution of
the corresponding first order conditions, which are given by
\[
R_{k}( c_0,...,c_T ) S^{i}_{k} = E \left[  R_{k + 1 }( c_0,...,c_T ) \left(  S^{i}_{k+1} + d^{i}_{k+1}
\right) | \mathcal{G}_k \right],
\]
and
\[
R_{k}( c_0,...,c_T ) = E \left[ R_{k + 1 }( c_0,...,c_T ) \left(  1 + r_{k+1} \right)
 |  \mathcal{G}_k \right],
\]
for all $i=0,...,N$ and $ k = 0 , ... ,  T-1$. Thereby, we
conclude that the process $ ( R_{k}( c^{*}_0,...,c^{*}_T ) )_{ k = 0 , . . . ,
T } $ is a SPD. Now, observe that $R_{T}( c^{*}_0,...,c^{*}_T ) =
u'_{T} ( c^{*} ( T , \beta , h )  )$ $ > 0$ by definition, and hence the
positivity of the process $ ( R_{k}( c^{*}_0,...,c^{*}_T ) )_{ k = 0 , . . . ,
T } $ follows by the above first order conditions and the fact
that $ (S_{k})_{ k = 0 , ... , T} , (d_{k})_{ k = 1
, ... , T} $ and $(r_k)_{ k = 1 , ... , T} $ are positive processes. 
Finally, the proof
is accomplished by applying Theorem \ref{spd} on $ (
R_{k}( c^{*}_0,...,c^{*}_T ) )_{ k = 0 , . . . , T } $ and combining this with the convention that
the aggregate SPD $ ( M_{k} )_{ k = 0 , . . . , T }$ does not vanish. $ \qed$
\newline
\newline
For the sake of convenience, the following notation will be used in the sequel.

\begin{Definition}
The process $ ( \widetilde{M}_k )_{k=0,...,T}$, given by
\begin{equation}
\widetilde{M}_{k} =  M_k +  \sum_{ l = k+1 }^{T} \
\sum_{j=1}^{l-k} \ \sum_{ k \leq s_{j} < ... < s_1 < l } \beta^{ (
l )  }_{s_1} \beta^{(s_1)}_{s_2} ... \beta^{(s_{j})}_{k} E \left[
M_l  \big| G_k \right] , 
\label{aggregate_spd}
\end{equation}
is referred as the \textit{aggregate perturbed SPD}.
\label{per_spd}
\end{Definition}

\section{Optimal Consumption for Some Special Classes of Financial Markets and Preferences}

The system of first order equations (\ref{first1}) appearing in Theorem \ref{existence} is highly non-linear and closed form solutions can only be obtained
in some special settings. In the current section, we study several such cases.
As it will be demonstrated in the subsequent 3 subsections, these closed form solutions provide a detailed 
description of the structure of the optimal consumption stream in the presence of habits, and allow 
illustrating certain important characteristics.

\subsection{Power Utility with No Random Endowment}

Consider an agent represented by a power utility $u_{k}(x)
= e^{- \rho k }  \frac{x^{1 - \gamma } }{ 1 - \gamma },$ for
$ k = 0 , ... , T$. Here, $\gamma$ and $\rho$ stand for the degree of
risk aversion and impatience respectively. Assume that the
agent receives only an initial endowment, that is, $\epsilon_0 \geq 0$, and $\epsilon_k = 0 $, for
all $ k = 1 , ... , T$. Assume further that no exogenous habits
are involved, namely, $h_k=0$, for all $ k = 1 , ... , T. $
We denote by $ c_k (\epsilon_0 ) $, $W_k (\epsilon_0)$ and $I_k
(\epsilon_0)$ the optimal consumption, wealth and investment
processes respectively, viewed as functions of the initial endowment
$\epsilon_0$. The solution to the corresponding utility
maximization problem is given explicitly in the following statement.
\begin{Theorem}
Under the above assumptions, the optimal consumption and investment policies are linear
increasing functions of the wealth, that is,
\begin{equation}
 c_{k}(\epsilon_0) = A_k  W_{k}(\epsilon_0)  \ \ \ ; \ \ \ c_{0}(\epsilon_0) = A_0  \epsilon_0,
\label{linear_consumption}
\end{equation}
and
\begin{equation}
I_{k} (\epsilon_0) = \left( 1 - A_k \right)  W_{k}(\epsilon_0) \ \ \ ; \ \ \ I_{0}(\epsilon_0) = \left( 1 - A_0 \right) \epsilon_0,
\label{linear_investment}
\end{equation}
for all $ k = 1 , ... ,  T,$ where $( A_k )_{ k = 0 , ... , T} $ is an adapted processes taking values in the interval $( 0 ,  1 ],$ which is given by
\[
A_k = \frac{ c_k ( 1)  }{ c_k ( 1 )  +   E \big[
\frac{M_{k+1}}{M_{k}}  W_{k+1}(1) | \mathcal{G}_k \big] }
.
\]
Moreover, the corresponding wealth process $( W_{k} ( \epsilon_0 ) )_{ k=1,...,T}  $ satisfies
\[
W_{ k+1 } ( \epsilon_0 ) = \frac{ W_{k+1} ( 1  ) }{ c_k (1)   }  c_k ( \epsilon_0 ),
\]
for all $k= 0 , ... , T-1$.
\label{det_endowments}
\end{Theorem}
\textbf{Proof of Theorem \ref{det_endowments}} First, observe
that the first order conditions (\ref{first1}) in Theorem
\ref{existence} imply that $ c_k (\epsilon_0) = c_{k}(1)
\epsilon_0 $ and $ W_k (\epsilon_0) =  W_{k}(1) \epsilon_0,$ for
all $k = 0 ,...,T$. This implies that $ W_{k+1}( \epsilon_0 ) =
\frac{W_{ k+1}(1)}{ c_{ k}(1)}  c_k( \epsilon_0 ) ,$ for all $ k =
0 , ... ,T-1,$ and by plugging this into equation (\ref{first2}),
we conclude the validity of (\ref{linear_consumption}). Next,
notice that relation (\ref{linear_investment}) holds due to Lemma
\ref{bc_parametrization}. We claim now that $ E \big[
\frac{M_{k+1}}{M_{k}}  W_{k+1}(1) | \mathcal{G}_k \big] > 0 , $
$P-$a.s., for all $ k= 0 , ... , T-1$. This is unclear a-priori
since the aggregate SPD $(M_k)_{k=0,...,T}$ is not necessarily
positive. To this end, set $\epsilon_0 =1$ and note that by
(\ref{first2}), we have $ W_{T}(1) = c_{T}(1) > 0 $. 
Therefore, by item (ii) in Remark (\ref{remark000}), we get
\[
E \left[ \frac{M_T }{ M_{T-1} } W_{T}(1)  \big| \mathcal{G}_{T-1} \right] = 
E \left[ \frac{ R_T }{ R_{T-1} } W_{T}(1)  \big| \mathcal{G}_{T-1} \right] > 0  ,
\]
where $ \left( R_{k} \right)_{k=0,...,T} $ is an arbitrary positive SPD.
Next, note that (\ref{bc_parametrization}) implies that $W_{T-1}(1) = c_{T-1}(1) + 
E \left[ \frac{M_T }{ M_{T-1} } W_{T}(1)  \big| \mathcal{G}_{T-1} \right]  > 0. $
Hence, as above, we conclude that $
E \left[ \frac{M_{T-1} }{ M_{T-2} } W_{T-1}(1)  \big| \mathcal{G}_{T-2} \right] = 
E \left[ \frac{ R_{T-1} }{ R_{T-2} } W_{T-1}(1)  \big| \mathcal{G}_{T-2} \right] > 0  .$
The rest of the claim can be proved analogously by using backward induction. $\qed$

\subsection{Exponential Utility and One Period Risk Free Bonds} 
Consider a
market which consists only of one period risk free bonds represented by an
interest rate process $(r_k)_{ k = 1 , ... , T} $. One checks that
$\mathcal{L}_{k} = L^{2} ( \mathcal{G}_{k-1}) $ (in particular, this
implies that $P_{\mathcal{L}}^{k} \left[  \cdot \right] = E \left[
\cdot | \mathcal{G}_{k-1} \right] $), and $M_{k} = \prod_{i=1}^{k}
(1+r_i)^{-1}$, for all $ k =1 , ... ,  T.$ The agent is
represented by an exponential utility $ u_{k} : {R} \to
{R}_{-} $, $ u_{k}(x) = - e^{ - \gamma x}$, $k=0,...,T,$
where $\gamma > 0$ stands for the degree of risk aversion.
We emphasize that this specification of preferences is
exceptional for the current paper, since the utility function is
defined on the whole real-line $ {R}$. Note however that the Inada
conditions hold in the following modified form, $ u'_k( - \infty
)= + \infty $ and $ u'_k( + \infty ) = 0,$ $k=0,...,T$. The
exogenous habits are characterized by the process $(h_k)_{k=0}$.
The internal habits coefficients are assumed to be homogeneous and the 
habit consumption level is assumed to depend only on 
last period consumption, i.e.,
$\beta^{(k)}_{k-1} = \beta \geq 0$, and $\beta^{(k)}_{l} = 0 $,
for all $ k = 1 , ... ,  T, $ and $ l = 0, ... , k-2.$ The latter
framework is in fact the most prevalent specification of a habit
forming utility function in the discrete time literature (see e.g. Abel \cite{A},
Gomes and Michaelides \cite{GM} and Polkovnichenko \cite{P}).
The utility maximization problem in the current setting is similar
to the one in (\ref{utility_max}), apart from the evident
distinction that the process $\left( c_k \right)_{k=0,...,T}$ is allowed to
take negative values. Let $ ( \epsilon_k )_{ k = 0 , ... , T} $ represent
the (possibly random) endowment stream of the agent. The optimal consumption, investment and wealth
are denoted respectively by $ (c^{*}_k)_{ k =0 , ... , T }$, $ ( I^{*}_k)_{ k =0 , ... , T }$ and
$ (W^{*}_k)_{ k = 1 , ... , T }$.

\begin{Theorem} Under the above assumptions, the optimal consumption and investment processes
are determined explicitly through the following recursive scheme,
\begin{equation}
c^{*}_k =  l_k  W^{*}_k  + m_{k} c^{*}_{k-1} + n_k( \epsilon_k , ... , \epsilon_T)  ,
\label{exp1} 
\end{equation}
\begin{equation}
I^{*}_k  =  \left(  1 - l_k \right) W^{*}_k - m_k c^{*}_{k-1} + \epsilon_k - n_k( \epsilon_k , ... , \epsilon_T) ,
\label{exp2} 
\end{equation}
\begin{equation}
W^{*}_k =  l'_k c^{*}_{k-1}  + m'_{ k} c^{*}_{k-2}  + n'_{k}( \epsilon_k , ... , \epsilon_T) ,
\label{exp3}
\end{equation}
for all $ k = 1 , ... , T $,
\[
c^{*}_0 =  l_0 \epsilon_0 + m_0 \frac{ n'_0 }{ m'_0 }   ,
\]
and
\[
I^{*}_0  = \epsilon_0 - c^{*}_0
\]
where $ l_k , m_k , n_k ( \epsilon_k , ... , \epsilon_T ), l'_k ,
m'_k$ and $n'_k ( \epsilon_k , ... , \epsilon_T ) $ are given
explicitly in (\ref{lm'}), (\ref{n'}), (\ref{lm}) and (\ref{n}). Furthermore, we have
\begin{equation}
0 < l_k \leq 1,
\label{exp4}
\end{equation}
for all $k=0,...,T$.
\label{exponential_utility}
\end{Theorem}

\textbf{Proof of Theorem \ref{exponential_utility}} As in the
proof of Theorem \ref{existence}, one checks that the optimal
consumption stream is determined as the unique solution of the
system of equations (\ref{first1}) and (\ref{first2}). Now, note
that equations $(\ref{first2})$ become $ c_{k} = \epsilon_k +
W_{k} - (1 + r_{k+1} )^{-1}  W_{k+1} ,$ for $k=0,...,T$. Thus
it is evident that $ l_T = 1 $, $ m_T = 0 $ and $n_T ( \epsilon_T ) = \epsilon_T.$
It is easy to check by backward induction that equations $ (\ref{first1}) $
admit the form
\[
E \left[ e^{ - \gamma \left( c_{k} - \beta c_{k-1}   - h_k \right) } X_k \big| \mathcal{G}_{k-1}  \right] = 
e^{ - \gamma \left( c_{k-1} - \beta c_{k-2}   - h_{ k-1 } \right) } ,
\]
for $k=1,...,T$, where $\left( X_k  \right)_{k=1,...,T}$ is given
by
\[
X_k = \beta + \frac{1 }{ \sum_{ j = k }^{T } \prod_{ i = j }^{ T }
\left( 1 + r_i  \right)^{-1} \beta^{ T - j }    }.
\]
Finally, the above formula combined with equations (\ref{first2}) implies that
\begin{equation}
l'_k = \frac{ 1 + \beta - m_k  }{ l_k }  \ \ \ ; \ \ \  m'_k = -
\frac{ \beta }{ l_k },
\label{lm'}
\end{equation}
\begin{equation}
n'_k ( \epsilon_k , ... , \epsilon_T  ) = \frac{ \log \left( E
\left[ \exp \left( \gamma \left( h_k -  n_k ( \epsilon_k , ... ,
\epsilon_T  ) \right)  \right) X_k  \big| \mathcal{G}_{k-1}
\right]  \right) - \gamma h_{k-1} }{ \gamma l_k }   ,
\label{n'}
\end{equation}
\begin{equation}
l_{k-1} = \frac{ 1 }{ 1 + ( 1 + r_{k} )^{-1} l'_k } \ \ \ ; \ \ \
m_{k-1} = - \frac{ ( 1 + r_{k} )^{-1} m'_k  }{1 + ( 1 + r_{k}
)^{-1} l'_k}  ,
\label{lm}
\end{equation}
\begin{equation}
n_{k-1} ( \epsilon_{k-1} , ... , \epsilon_T  )  = \frac{
\epsilon_{k-1} - ( 1 + r_{k} )^{-1} n'_k ( \epsilon_{k} , ... ,
\epsilon_T  )   }{ 1 + ( 1 + r_{k} )^{-1} l'_k },
\label{n}
\end{equation}
for $k=1,...,T$. Finally, one checks by induction that $l_k \in [0,1),$ $k=0,...,T$. $\qed$

\subsection{Complete Markets}
By definition, the relation $ L_{k} = L^{2}( \mathcal{G}_k) $ (and in particular,
$P_{\mathcal{L}}^{k} \left[  \cdot \right] = E \left[  \cdot  | \mathcal{G}_k \right]$) holds for all $ k = 1 , ... , T,$ in a complete market setting.
Moreover, the aggregate SPD is the unique positive SPD in the market.
The next statement treats the case of general habit-forming utility functions.

\begin{Theorem}
The optimal consumption stream $ \left( c^{*}_k   \right)_{ k = 0 , ... , T}  $ is determined explicitly through the following recursive scheme,
\begin{equation}
 c^{*}_{k} = \sum_{l=0}^{k-1} \beta^{(k)}_{l} c^{*}_l + h_k + \left( u'_{k} \right)^{-1} \left( u'_{ k-1 } \left( c^{*} ( k-1 , \beta , h_{ k-1 }    )    \right)
  \frac{ \widetilde{M}_{k} }{ \widetilde{M}_{k-1}} \right),
\label{complete1}
\end{equation}
for all $ k =1 , ... ,  T $, and
\begin{equation}
 \sum_{  k = 0 }^{ T } E \left[  c^{*}_k  \frac{M_{k}}{M_{k-1}}  \right] = \sum_{ k = 0 }^{ T} E \left[  \epsilon_k  \frac{M_{k}}{M_{k-1}}
 \right],
\label{complete2}
\end{equation}
where $ \big( \widetilde{M}_k \big)_{k=0,...,T}$ is given in (\ref{aggregate_spd}), and
$( c^{*} ( k , \beta , h    ) )_{k=0,...,T} $ is defined as in
(\ref{per_spd}).
\label{complete_gen_ut}
\end{Theorem}

\textbf{Proof of Theorem \ref{complete_gen_ut}} Since the market
is complete, the budget constraint equations $(\ref{first2})$ are
redundant for $ k =1 , ... ,T$ and thus the only relevant budget
constraint is equation (\ref{first2}) for $k=0.$ This equation can
be rewritten as (\ref{complete2}) by identity (\ref{budgetc2}) in
Lemma \ref{bc_parametrization}. The first order conditions
(\ref{first1}) can be transformed into (\ref{complete1}) by
combining the identity $P_{\mathcal{L}}^{k} \big[  \cdot \big] =
E \big[  \cdot | \mathcal{G}_k \big] $, $ k = 1 , ... ,  T,$
with some routine, but somewhat tedious, algebraic manipulations (similarly as in Lemma \ref{reformulation}). \qed
${}$
\newline
We concentrate now on power utility functions. That is, $u_{k}( x ) =   e^{- \rho k } \frac{ x^{1- \gamma_k } }{
1- \gamma_k } $, for $k = 0 , ... , T$, where $ ( \gamma_k )_{ k =
0 , ... , T}  $ is a sequence of non-negative numbers that
represent the risk aversion of the agent, and $\rho$ indicates the impatience coefficient.
In contrast to the previous two subsections, the
optimal consumption stream here may demonstrate a non-linear
structure. The next result illustrates the latter phenomenon, and presents
an analytical solution to the associated utility maximization problem.

\begin{Theorem} (i) Under the above assumptions, we have
\[
c^{*}_k  = \psi_k \left(  W^{*}_k  \right),
\]
for $k=1,...,T$ , and $ c^{*}_0  = \psi_0 \left(  \epsilon_0   \right)$ where, $ \psi_k : L^2( \mathcal{G}_k) \to L^2( \mathcal{G}_k) $ is a
smooth function which is
given explicitly by (\ref{H}), for all $k=0,...,T$. The derivative $ d \psi_k $ is a
multiplication operator such that $ 0 < (d \psi_k )_{ W_k( \epsilon_0 ,
... , \epsilon_{k-1} ) } $ $ \leq 1.$ Furthermore, we have
\[
I_k( \epsilon_0 , ... , \epsilon_{k-1} )  = \widehat{ \psi }_k \left(
W_k( \epsilon_0 , ... , \epsilon_{k-1} )  \right),
\]
for $k=1,...,T$ and $ I_{0} ( \epsilon_0 ) = \widehat{ \psi }_k \left( \epsilon_0  \right) $.
Here, $ \widehat{ \psi }_k :  L^2( \mathcal{G}_k) \to L^2(
\mathcal{G}_k) $ is given by $ \widehat{ \psi }_k (X) = \epsilon_k +  X
-  \psi_k (X) $, for all $k=0,...,T$. In particular, the optimal consumption/investment is
an increasing function of the wealth.
\newline
(ii) If $ \gamma_0 = ... = \gamma_T , $ then, the optimal consumption/investment stream is a linear increasing
function of the wealth. More precisely,
\[
\psi_{k}(X) = \frac{  \sum_{i=  0 }^{k} D^{(i)}_k }{ \sum_{i=0  }^{T}
F^{(i)}_k } X + \frac{  \sum_{i= 0 }^{k} D^{(i)}_k }{ \sum_{ i = 0
}^{T} F^{(i)}_k } \sum_{j=k}^{T} E \left[ \frac{M_j}{M_k}
\epsilon_j   \big| \mathcal{G}_k  \right] + \sum_{ i =1 }^{ k }
\alpha^{(k)}_i h_i,
\]
where $D^{ ( i ) }_{k}$, $\alpha^{(k)}_i$ and $ F^{(i)}_k $ are
given by (\ref{D}), (\ref{alpha}) and (\ref{F}) respectively.
\label{complete_power}
\end{Theorem}
\textbf{Proof of Theorem \ref{complete_power}}
(i) Theorem \ref{complete_gen_ut} implies that
$ ( c^{*}_k )_{ k = 0 , ... , T}$ satisfies the equation
\[
c^{*}_k  - \sum_{ j = 0 }^{ k-1 } \beta^{(k)}_j c^{*}_j  - h_k  = ( c^{*}_0 ) ^{ \gamma_0 / \gamma_k } e^{- \frac{ \rho k }{ \gamma_k } } (\widetilde{M}_k )^{- 1 / \gamma_k },
\]
for $k=1,...,T$. By using routine algebraic operations, one obtains that
\begin{equation}
c^{*}_{k}  =  \sum_{ i = 0 }^{k} D^{ ( i ) }_{k}  (c^{*}_0)^{  \gamma_0 / \gamma_i } + \sum_{ i = 0 }^{ k } \alpha^{(k)}_i h_i,
\label{consumption_power}
\end{equation}
for $k=1,...,T$, where
\begin{equation}
D^{ ( i ) }_{k} = \delta^{( i )}_k e^{ - \frac{ \rho i }{ \gamma_i } }  \left( \widetilde{M}_i \right)^{ -  1 / \gamma_i  } .
\label{D}
\end{equation}
The processes  $( \delta^{( i )}_k )_{ k=1 , ... ,T ; i =1 , .... , k } $ and $ ( \alpha^{( i )}_k )_{ k=1 , ... ,T ; i =1 , .... , k } $ are determined
through the following recursive scheme: $ \delta^{(0)}_0 = 1 $, $ \delta^{(0)}_1 = \beta^{(1)}_0,$ $ \delta^{(1)}_1=1,$ $ \alpha^{(1)}_1=1,$ and
\begin{equation}
\alpha^{(k)}_i = \sum_{ j = i }^{ k - 1 } \beta^{(k)}_j \alpha^{ (
j ) }_i    \ \ \ ; \ \ \  \alpha^{(k)}_{k} =  1 , \label{alpha}
\end{equation}
\begin{equation}
\delta^{ ( i ) }_k =  \sum_{ j = i }^{ k-1 }  \beta^{(k)}_{j} \delta^{(i)}_j \ \ \ ; \ \ \ \delta^{( k )}_k = 1 ,
\label{delta}
\end{equation}
for all $k= 2 ,...,T$ and $i= 0 , ... , k-1$. Next, identity (\ref{budgetc2}) in Lemma \ref{bc_parametrization} yields
\begin{equation}
W^{* }_k  = \sum_{i= 0 }^{T} F^{(i)}_k  (c^{* }_0)^{ \gamma_0 / \gamma_i }  - \sum_{j=k}^{T} E \left[ \frac{ M_j }{ M_k }  \epsilon_j  \big| \mathcal{G}_k  \right] ,
\label{wealth_power}
\end{equation}
where
\begin{equation}
F^{(i)}_k = \sum_{j= \max ( i , k ) }^{T} E \left[ \frac{ M_j}{ M_k } D^{(i)}_{j} \big| \mathcal{G}_k \right] ,
\label{F}
\end{equation}
for all $ k = 1,...,T$ and $ i= 0,...,T.$ Equations
(\ref{consumption_power}) and (\ref{wealth_power}) allow to
establish a correspondence between the optimal consumption and the
wealth as follows. Let $f_{k} : L^{2}_{+} ( \mathcal{G}_k ) \to
L^{2} ( \mathcal{G}_k ) $ be defined by $ f_{k} ( X ) = \sum_{ i =
0 }^{T} F^{(i)}_{k} X^{ \frac{ \gamma_0 }{ \gamma_i } }  -
\sum_{j=k}^{T} E \left[ \frac{ M_j }{ M_k }  \epsilon_j  \big|
\mathcal{G}_k  \right] $. Since each function of the form $ x
\mapsto x^{ \eta } $ is increasing, for an arbitrary $ \eta \geq 0,$ it
follows that $f_k$ is an invertible function. Let us denote by
$g_k$ the inverse of $f_k$, which is defined on $Im( f_k )
\subseteq L^{2}( \mathcal{G}_k )$. Next, consider the function
\begin{equation}
\psi_k(X) = \sum_{ i = 0 }^{k} D^{(i)}_{k}  g^{ \frac{ \gamma_0 }{
\gamma_i } }_k ( X )  + \sum_{ i =  0 }^{ k} \alpha^{(k)}_i h_i,
\label{H}
\end{equation}
and note that $c^{*}_k = \psi_k (
W^{*}_k ). $ Observe that $ (d
f_k)_{X} Y = \sum_{ i = 0 }^{T} $ $F^{(i)}_k \frac{ \gamma_0 }{
\gamma_i } X^{ \frac{ \gamma_0 }{ \gamma_i } -1  } Y, $ hence,
$ (d f_k)_{X} = \sum_{ i = 0 }^{T} F^{(i)}_k \frac{ \gamma_0 }{
\gamma_i } X^{ \frac{ \gamma_0 }{ \gamma_i } -1  } $ is a
multiplication operator. Therefore, we conclude that
\[
(d g_k)_{X}Y  = \left( \sum_{ i = 0 }^{T} F^{(i)}_k \frac{ \gamma_0 }{ \gamma_i } ( g_k(X) )^{ \frac{ \gamma_0 }{ \gamma_i } -1  } \right)^{-1} Y
.\]
The definition of $ \psi_k$ implies that
\[
(d  \psi_k )_{X} Y = \sum_{i=0}^{k} D^{(k)}_i \frac{ \gamma_0 }{
\gamma_i } ( g_k (X))^{ \gamma_0 / \gamma_i - 1} (d g_k)_{X} Y,
\]
which implies in particular that $ (d  \psi_k )_{X} $ is a
multiplication operator. It is left to check that $ 0 < (d \psi_k
)_{ W^{* }_k } \leq 1.$ To this end, note that
\[
(d \psi_k )_{ W^{* }_k } = \frac{  \sum_{i= 0 }^{k} D^{(i)}_k \frac{ \gamma_0 }{ \gamma_i } \left( c^{* }_0  \right)^{ \frac{ \gamma_0 }{ \gamma_i } -1 }  }{ \sum_{ i = 0 }^{T} F^{(i)}_k \frac{ \gamma_0 }{ \gamma_i } \left( c^{* }_0  \right)^{ \frac{ \gamma_0 }{ \gamma_i } -1 }  },
\]
and the claim follows since $ D^{(i)}_k \leq F^{(i)}_k, $ for all $i=0,...,T$ and $k=0,...,T.$
Finally, observe that equations (\ref{H}) and (\ref{complete2}) imply that
\[
\psi_0( \epsilon_0 )  + \sum_{i= 0 }^{T} \sum_{k=i}^{T} E \left[ M_k D^{(i)}_k \right] \psi^{ \gamma_0 / \gamma_i }_{0}( \epsilon_0 )  - \epsilon_0 - \sum_{k=1}^{T} \sum_{i=1}^{k} E \left[ M_k  \alpha^{(k)}_i h_i  \right] = 0 ,
\]
and thus by differentiating this equation with respect to $\epsilon_0$, we obtain that
\[
0 < \psi_0'( \epsilon_0 ) = \frac{1}{
 1  + \sum_{i=1}^{T} \sum_{k=i}^{T} E \left[ M_k D^{(i)}_k \right] \psi^{ \gamma_0 / \gamma_i -1 }_{0} ( \epsilon_0 ) }  <  1.
\]
(ii) The assertion follows easily by part (i). $\qed$

\section{General Incomplete Markets: Explicit Solution and Monotonicity}

In this section we study in more detail an individual's habit-forming utility maximization problem in
a general setting of incomplete markets. This general framework requires a more delicate investigation than the one
exhibited in the previous section, due to the presence of random endowments,
additive habits and general preferences. We will provide an analytic solution
to the utility maximization problem through an explicit recursive construction based on the implicit
function theorem. We will use this construction to investigate monotonicity properties of the optimal
consumption/investment policy. 

\subsection{Dynamic Programming Fails in Showing Monotonicity}

A conventional approach to tackle the monotonicity problem would be
by using dynamic programming (see \cite{PL}). Despite that this general methodology
provides a useful characterization of the controls involved in
the optimization problem, it cannot be directly applied to proving monotonicity in the current setting. We will illustrate this below for the problem of detecting the dependence of the initial consumption $c_0 ( \epsilon_0 ) $ on the initial endowment $\epsilon_0.$  
\newline
Consider the value function associated with
the utility maximization problem (\ref{utility_max}) viewed as a function of
$ \epsilon_0$:
\begin{equation}
V_{0}( \epsilon_0) :=  \sup_{ \pi \in \Pi_0 } \ \sum_{k=0}^{T} E \big[ u_{k} \big(  c( k , \beta , h , \pi ) \big) \big].
\label{AAA}
\end{equation}
Here, $ c( k , \beta , h , \pi ) := c^{\pi}_{k} - \sum_{ j =
0 }^{k-1} \beta^{(k)}_{j} c^{\pi}_{j} - h_k $; $ c^{\pi}_k$ is defined analogously to (\ref{consumption});
and $\Pi_0$ denotes the set of all trading strategies $\pi=(\pi_0,...,\pi_{T-1},0)$ such that
$ c( k , \beta , h , \pi ) \geq 0$, for all
$k=0,...,T$. Note that $c^{\pi}_0$ depends on $\epsilon_0$, and thus each
$ c( k , \beta , h , \pi ) $ depends on $\epsilon_0$, unless $ \beta^{(k)}_0=0.$
Similarly, for each $k=1,...,T$, and fixed portfolio (up to the period $k-1$) $\widehat{\pi}_o, ... , \widehat{\pi}_{k-1},$
we consider the $\mathcal{G}_k-$measurable random variable:
\[
V_k ( \epsilon_0 , \widehat{\pi}_o , ... , \widehat{\pi}_{k-1} ) = 
\]
\[
\sup_{ \pi \in \Pi ( \widehat{\pi}_o , ... , \widehat{\pi}_{k-1} ) }
u_{k} \left( c( k , \beta , h , \pi )  \right) + \sum_{ j = k+1}^{T} E \left[  u_{j} \left(
c( j , \beta , h , \pi ) \right) | \mathcal{G}_k \right],
\]
where $ \Pi ( \widehat{\pi}_o , ... , \widehat{\pi}_{k-1} ) \subseteq \Pi_0 $ denotes the set of all trading strategies
of the form $(\widehat{\pi}_o , ... , \widehat{\pi}_{k-1}, \pi_k , ... , \pi_{T-1}, 0)$. Since at the maturity date we have $\pi_T = 0$,
it follows that
\[
 V_{T} ( \epsilon_0 , \widehat{\pi}_0 , ... , \widehat{\pi}_{T-1} ) = u \left( c \left( T , \beta , h , \widehat{\pi} \right)  \right),
\]
where $ \widehat{\pi} = ( \widehat{\pi}_0 , ... ,
\widehat{\pi}_{T-1} , 0 ).$ Let us assume for simplicity that the
market excludes redundant assets. This implies the  existence of a
unique optimal portfolio $ \pi(\epsilon_0) = ( \pi_0 (\epsilon_0)
, ... , \pi_T (\epsilon_0) )  $. Furthermore, it follows (as in
subsection 4.2) that each $\pi_{k}( \epsilon_0) $ is component-wise
differentiable with respect to $\epsilon_0$. We denote by $ \pi'_k
(\epsilon_0 ) $ the corresponding gradient. Now, recall that $ c(0
, \beta, h, \pi ) = \epsilon_0 - \pi_0 \cdot S_0 ,$ hence, the
dynamic programming principle implies that
\[
 V_{0}( \epsilon_0) =  \sup_{ \pi_0  ,  \Pi ( \pi_0 ) \neq \emptyset  } \left\{ u \left( \epsilon_0 - \pi_0 \cdot S_0 \right) + E \left[
V_{1} \left( \epsilon_{0} , \pi_{0} \right)  \right] \right\}
\]
\[
= u \left( \epsilon_0 - \pi_0 (\epsilon_0) \cdot S_0 \right) + E \left[
V_{1} \left( \epsilon_{0} , \pi_{0} (\epsilon_0) \right)  \right] .
\]
Similar arguments yield 
\[
 V_{k} \left( \epsilon_0 , \widehat{\pi}_0 , ..., \widehat{\pi}_{k-1} \right) =
\sup_{ \pi_{k} , \Pi ( \widehat{\pi}_0 , ..., \widehat{\pi}_{k-1} , \pi_k) \neq \emptyset  }  \bigg( u_k \left( c \left( k , \beta, h , \pi \right)  \right)
-
\]
\[
E \left[ V_{k} \left( \epsilon_{0} , \widehat{\pi}_0 ,... , \widehat{\pi}_{k-1}  , \pi_{k} ( \epsilon_0 )  \right) \big| \mathcal{G}_k  \right] \bigg) ,
\]
and
\[
 V_{k} \left( \epsilon_0 , \pi_0 ( \epsilon_0 ) , ..., \pi_{k-1} ( \epsilon_0 ) \right) =
u_k \left( c (k , \beta, h , \pi( \epsilon_0)  )  \right) +
\]
\[
E \left[ V_{k} \left( \epsilon_{0} ,  \pi_{0} ( \epsilon_0 ) ,... ,  \pi_{k-1} ( \epsilon_0 )  , \pi_{k} ( \epsilon_0 )  \right) \big| \mathcal{G}_k  \right],
\]
for all $k=1,...,T.$ Observe that $  V_{T} ( \epsilon_0 ,
\widehat{\pi}_0 , ... , \widehat{\pi}_{T-1} ) $ is differentiable
with respect to each coordinate, by definition. Therefore, the
above recursive relation combined with the differentiability of
$\pi( \epsilon_0)$ implies that each $ V_{k} \left( \epsilon_0 ,
\pi_0 , ..., \pi_{k-1} \right) $ is differentiable with respect to
all coordinates. We claim that
\begin{equation}
 V'_0(\epsilon_0) = u'_{0}( c_0 (\epsilon_0) ) - \sum_{ k=1 }^{T} \beta^{(k)}_{0} E \big[ u'_{k}( c ( k , \beta, h , \pi ( \epsilon_0 ) )  ) \big],
\label{DPE}
\end{equation}
where $( c_k ( \epsilon_0 ) )_{k=0, ... ,T}$ denotes the optimal consumption stream. To this end, differentiate equation (\ref{aaa}) with respect to $\epsilon_0$,
and observe that
\[
 V'_{0} ( \epsilon_0 ) = u'(c_0 ( \epsilon_0 ) )  \left( 1 - \pi'_{0}( \epsilon_0 ) \cdot S_0 \right) +
\]
\[
E \left[ \frac{ \partial V_1 }{ \partial \epsilon_0 }  \left( \epsilon_{0} , \pi_{0} (\epsilon_0) \right) \right] +
 E \left[ \frac{ \partial V_1 }{ \partial \pi_0 } \left( \epsilon_{0} , \pi_{0} (\epsilon_0) \right) \cdot \pi'_{0} (\epsilon_0) \right],
\]
where
\[
\frac{ \partial V_1 }{ \partial \pi_0 } \left( \epsilon_{0} , \pi_{0} (\epsilon_0) \right) =
\left( \frac{ \partial V_1 }{ \partial \phi_0 } \left( \epsilon_{0} , \pi_{0} (\epsilon_0) \right) , \frac{ \partial V_1 }{ \partial \pi^{1}_0 } \left( \epsilon_{0} , \pi_{0} (\epsilon_0) \right) , ... , \frac{ \partial V_1 }{ \partial \pi^{N}_0 } \left( \epsilon_{0} , \pi_{0} (\epsilon_0) \right) \right).
\]
Since $\pi_0( \epsilon_0 ) $ is the optimal control, the first-order conditions imply that the following equation is satisfied (component-wise):

\[
  u'_0 \left( \epsilon_0 - \pi_0 ( \epsilon_0 ) \cdot S_0 \right) S_0   =
  E \left[ \frac{ \partial V_1 }{ \partial  \epsilon_0 }  \left(  \epsilon_0 , \pi_0 ( \epsilon_0 ) \right)
  \right].
\]
Therefore, we get
\[
 V'_0 \left( \epsilon_0 \right) = u'(c_0 ( \epsilon_0 ) ) + E \left[  \frac{ \partial V_1 }{ \partial \epsilon_0 }
 ( \epsilon_0 , \pi_0 ( \epsilon_0 ) ) \right].
\]
We turn now to computing the random variable $\frac{ \partial V_1 }{
\partial \epsilon_0 }  ( \epsilon_0 , \pi_0 ( \epsilon_0 ) ) $ explicitly. As before, one checks that
\begin{equation}
 \frac{ \partial V_{k} }{ \partial \epsilon_0 } ( \epsilon_0 , \pi_0 (\epsilon_0) , ... , \pi_{k-1} (\epsilon_0) ) =  - \beta^{(k)}_{0} u'
( c( k , \beta, h , \pi (\epsilon_0) )  )  +
\label{aaa}
\end{equation}
\[
E \left[  \frac{ \partial V_{k+1} }{ \partial \epsilon_0 }  ( \epsilon_0 , \pi_0 (\epsilon_0) , ... , \pi_{k-1} (\epsilon_0) , \pi_k( \epsilon_0) ) ) | \mathcal{G}_{k}  \right],
\]
for all $k=0,...,T-1.$ By definition, we have
\[
 \frac{ \partial V_T }{ \partial \epsilon_0 } ( \epsilon_0 , \pi_0 (\epsilon_0) , ... , \pi_{T-1} (\epsilon_0)  = - \beta^{(T)}_{0} u'_T \left(
c ( T , \beta, h , \pi( \epsilon_0 ) )  \right).
\]
Backward induction implies that
\[
 E \left[  \frac{ \partial V_1 }{ \partial \epsilon_0 } ( \epsilon_0   , \pi_0 ( \epsilon_0 ) ) \right] = - \sum_{k=1}^{T} \beta^{(k)}_{0} E \big[ u'_{k} ( c (k , \beta, h , \pi ( \epsilon_0 )  )  )
 \big].\]
This proves the validity of (\ref{DPE}). Assume that there are \textit{no habits}, i.e., $\beta^{ (k  ) }_{j} =0$, and $h_k=0$, for all $k=0,...,T$ and
$j=0 , ... , k-1.$ Then, (\ref{DPE}) takes the form
\[
V'_0(\epsilon_0) = u'_{0}( c_0(\epsilon_0)).
\]
Note that the value function $V_0(\epsilon_0)$ is concave, since $ u_{k} $ is concave, for all $k=0,...,T$, and thus we obtain that
$c'_0(\epsilon_0) =\frac{V_0''(\epsilon_0)}{u_0''(c_0(\epsilon_0))}\geq 0 $. 
However, in the presence of habits, it does not seem feasible to use the (much more complex) dynamic programming principle (\ref{DPE}) to derive the required  monotonicity of $c_0(\epsilon_0).$ Namely, the  presence of the cumbersome term 
$$ - \sum_{ k=1 }^{T} \beta^{(k)}_{0} E \big[ u'_{k}( c ( k , \beta, h , \pi ( \epsilon_0 ) )  ) \big] ,$$ 
makes the analysis significantly more complicated and other methods have to be developed. 
Note also that the other desirable inequality $c'_0(\epsilon_0) \leq 1 $ does \textbf{not} follow from (\ref{DPE}) even in the case without habits.

\subsection{Explicit Solution and Monotonicity}

We provide an explicit solution to the utility maximization
problem and study the associated monotonicity issues of the optimal
consumption/investment policy, in the setting of \textit{arbitrary} incomplete markets with a
deterministic interest rate, and for \textit{arbitrary}
idiosyncratically (incomplete) markets. We begin with a concise
description of the later type of markets. For some related works
involving the concept of idiosyncratic source of risk, the reader is addressed to
Duffie et al. \cite{DFSZ}, Hendersen \cite{H}, Malamud \cite{M} and Malamud and Trubowitz \cite{MT}.

\begin{Definition}
A financial market is called idiosyncratically incomplete, if there
exist two filtrations $( \mathcal{F}_k )_{ k = 0 , ... ,  T}$ and
$( \mathcal{G}_k )_{ k = 0 , ... ,  T}$ such that:
\newline
(i) $ \mathcal{F}_0 = \mathcal{G}_0 = \{ \emptyset , \Omega \}$, and
$ \mathcal{F}_k \subseteq \mathcal{G}_k $, for all $ k = 1 , ... , T.$
\newline
(ii) The market is complete with respect to $\mathcal{F}$, and the endowment
stream $( \epsilon_k )_{ k = 0 , ... , T } $ is adapted with respect to $\mathcal{G}.$
\newline
(iii) For each $k=0,...,T-1,$ and an arbitrary random variable $ X \in L^{2} \left( \mathcal{F}_{ k + 1 } \right) $, we have
\[
E \left[ X | \mathcal{G}_k \right] = E \left[ X | \mathcal{F}_k \right].
\]
\label{def1}
\end{Definition}

\begin{Lemma} For an arbitrary idiosyncratically incomplete market, we have
\newline
(i) For each $k=1,...,T$, the financial wealth space is given by
\[
\mathcal{L}_k = L^{2} \left( \sigma \left( \mathcal{G}_{k-1} , \mathcal{F}_{k}  \right) \right),
\]
and, in particular, $P^{\mathcal{L}}_k \left[ \cdot  \right] = E \left[ \cdot | \sigma \left( \mathcal{G}_{k-1} , \mathcal{F}_{k}  \right) \right].$
\newline
(ii) The aggregate SPD is a positive process, adapted with respect to $\mathcal{F}$.
Moreover, the aggregate SPD coincides with the unique positive SPD that
represents the same financial market with the filtration $\mathcal{F}$.
\label{lem1}
\end{Lemma}

\textbf{Proof of Lemma \ref{lem1}} See the proof of Propositions 4.2 and 4.3 in Malamud and Trubowitz \cite{MT}. \qed
${}$
\newline
The main result of this section heavily relies on the following important algebraic identities, 
leading to a severe simplification of the first order equations (\ref{first1}).

\begin{Lemma} For a given incomplete market with a deterministic interest rate, or for an idiosyncratically incomplete market,
the first order conditions (\ref{first1}) are equivalent to
\begin{equation}
P_{ \mathcal{L} }^{k} \left[ u'_k \left( c(  k , \beta , h_k ) \right) \right] = \frac{ \widetilde{M}_k }{ \widetilde{M}_{kt-1} }
 u'_{k-1} \left( c(  k-1 , \beta , h_{k-1} ) \right),
\label{eq1}
\end{equation}
for all $ k = 1 , ... ,  T,$ where the aggregate perturbed SPD $(\widetilde{M}_{k})_{k=1, ... ,T}$ is
defined in (\ref{aggregate_spd}).
\label{reformulation}
\end{Lemma}

\textbf{Proof of Lemma \ref{reformulation}} The proof is based on
the following simple observations: For markets with a deterministic
interest rate, one checks that the quantities $  E [ \frac{M_k}{M_{k-1}} |
\mathcal{G}_{k-1} ] $ and $  E [ \frac{ \widetilde{M}_k}{
\widetilde{M}_{k-1}} | \mathcal{G}_{k-1} ] $, are positive and
deterministic, for all $k=1,...,T$; for idiosyncratically incomplete markets, the
aggregate perturbed SPD satisfies: $ \widetilde{M}_{k} \in
\mathcal{F}_k$, for all $k=1,...,T$. We provide a unified proof
for both cases. For the sake of brevity, we set $
\widehat{c}_{k} :=  c(  k , \beta , h_k )$, for $k=0,...,T$.
Consider equation (\ref{first1}) for $k=T$:
\begin{equation}
P^{T}_{ \mathcal{L} } \left[ u'_T ( \widehat{c}_{T} )  \right]  = \frac{ M_T }{ M_{T-1} } \bigg( u'_{T-1} (  \widehat{c}_{T-1}  )  -
 \beta^{(T)}_{T-1} E \left[  u'_T ( \widehat{c}_{T}  ) | \mathcal{G}_{T-1} \right]  \bigg).
\label{eq2}
\end{equation}
An application of the conditional expectation $E [ \cdot | \mathcal{G}_{T-1} ] $ to both sides of this equation yields
\[
E \left[  u'_T ( \widehat{c}_{T} ) | \mathcal{G}_{T-1} \right] = \frac{ E \left[ M_T | \mathcal{ G}_{T-1}
\right] }{ M_{T-1} + \beta^{(T)}_{T-1} E \left[ M_T | \mathcal{ G}_{T-1}
\right] } u'_{T-1} ( \widehat{c}_{T-1} ) ,
\]
and by plugging this back into (\ref{eq2}), one concludes that
$(\ref{eq1})$ holds for $k=T$. Next, in the idiosyncratically incomplete case,
one can combine the preceding formula with property (i) in Lemma
\ref{lem1} (in the deterministic interest rate case, one can
exploit the fact that $ \frac{ M_{T-1}  }{ M_{T-1} +
\beta^{(T)}_{T-1} E \left[ M_T | \mathcal{ G}_{T-1} \right] }$ is
a positive constant), to check that equation (\ref{first1})
admits the following form, for $ k =T-1,$
\[
\frac{ M_{T-1}  }{ M_{T-1} + \beta^{(T)}_{T-1} E \left[ M_T | \mathcal{ G}_{T-1}
\right] }  P_{ \mathcal{L} }^{T-1} \left[ u'_{T-1} \left( \widehat{c}_{T-1}  \right) \right]  =
\]
\[
\frac{ M_{T-1} }{ M_{T-2} } \left( u'_{T-2} ( \widehat{c}_{T-2}  ) -
E \left[ \left( \beta^{(T-1)}_{T-2} + \beta^{(T)}_{T-2} \frac{ \widetilde{M}_T }{ \widetilde{M}_{T-1} }
 \right) u'_{T-1} ( \widehat{c}_{T-1}  ) | \mathcal{G}_{T-1}    \right]    \right) .
\]
Now, one can multiply the above equation by
\[
E \left[ \beta^{(T-1)}_{T-2} + \beta^{(T)}_{T-2} \frac{ \widetilde{M}_T }{ \widetilde{M}_{T-1} }
| \mathcal{G}_{T-1}    \right]  \frac{ M_{T-1} + \beta^{(T)}_{T-1} E \left[ M_T | \mathcal{ G}_{T-1}
\right] }{ M_{T-1}  } ,
\]
apply then the conditional expectation $E \left[ \cdot \big| \mathcal{G}_{T-2} \right] $ to both sides of the resulted equation,
and proceed as follows to conclude the validity of (\ref{eq1}), for $k=T-1$:
In the idiosyncratically incomplete case, one can use property (iii) in Definition \ref{def1} and the identity $ X P^{k}_{L} [ Y  ] = P^{k}_{L} [ X Y  ] ,$
which holds for all $Y \in L^{2} ( \mathcal{G}_T) $, $X \in L^{2} ( \mathcal{F}_{k} )$, and all $k=1,...,T$; in the deterministic interest rate case,
one can use the fact that $ E \left[ \beta^{(T-1)}_{T-2} + \beta^{(T)}_{T-2} \frac{ \widetilde{M}_T }{ \widetilde{M}_{T-1} }  | \mathcal{G}_{T-1}    \right]  $
and $ \frac{ M_{T-1} + \beta^{(T)}_{T-1} E \left[ M_T | \mathcal{ G}_{T-1} \right] }{ M_{T-1}  } $ are positive constants. The rest of the proof
follows by similar arguments and induction. $\qed$
\newline
\newline
Now, we present the main result of this section.

\begin{Theorem} Consider the utility maximization problem (\ref{utility_max}) in the setting
of an incomplete market with a deterministic interest rate, or an idiosyncratically incomplete market. Then, we have
\[
c^{*}_T = \psi_T \left( W^{*}_T  \right) ,
\]
\[
c^{*}_k = \psi_k \left( W^{*}_k  ,  c^{*}_{k-1} , ... , c^{*}_0  \right) ,
\]
for $k=1,...,T-1$, and
\[
 c^{*}_0  = \psi_0 \left( \epsilon_0 \right) ,
\]
\[
W^{*}_k  = \eta_k \left(  c^{*}_{k-1}  , ... , c^{*}_0  \right) ,
\]
for $k=2,...,T$, and
\[
W^{*}_1  = \eta_1 \left( c^{*}_0 \right) ,
\]
where the functions $ \psi_{T} $, $ \left( \psi_{k} \right)_{k=0,...,T-1} $ and $ \left( \eta_{k} \right)_{k=1,...,T} $
are given in (\ref{tec0}), (\ref{expl1}), (\ref{expl3}) and (\ref{expl5}). Furthermore, a monotonicity
property of the optimal consumption holds in the following form: The differential
of $\psi_{k}$, for $k=1,...,T$, with respect to $W_k$ is a multiplication
operator that satisfies:
\[
0 < \frac{ \partial \psi_k }{ \partial W_k } \leq \frac{1}{ 1 + \sum_{j=k+1}^{ T }
\left( \sum_{i=k}^{j} \beta_{k}^{ i } \beta_{i}^{ i+1 } ... \beta_{j-1}^{j}   \right) E \left[ \frac{ M_j }{ M_k }  \big| \mathcal{G}_k \right]  } ,
\]
for $k=1,...,T,$ and
\[
0 < \psi'_0 ( \epsilon_0 ) \leq   \frac{1}{ 1 + \sum_{j=1}^{ T }
\left( \sum_{i=0}^{j} \beta_{0}^{ i } \beta_{i}^{ i+1 } ... \beta_{j-1}^{j}   \right) E \left[  M_j \big| \mathcal{G}_k \right]  } .
\]
\label{main1}
\end{Theorem}
\textbf{Remark.}
As in the proof of Lemma \ref{reformulation}, we present a unified approach that treats both markets with
a deterministic interest rate and idiosyncratically incomplete markets. As one can notice, the additive habits structure in our model generate an essential distinction between the interaction of the optimal
consumption/investment rules in the last two periods. The reason for this comes from the fact that the consumption at the
maturity has no impact on future consumption. Therefore, we focus on proving the statement for the periods
$k=T-2, T-1, T$, which constitutes the core of Theorem \ref{main1}. The rest is then concisely explained, and can be proved
by induction, without any further machinery.
\newline
\newline
\textbf{Proof of Theorem \ref{main1}}
To simplify notations, we set
$
\widehat{c}^{*}_k  :=  c^{*}_{k}  -
\sum_{l=0}^{k-1} \beta^{(k)}_l c^{*}_l - h_k ,
$
for all $k=0,...,T$. For an arbitrary consumption stream $\left( c_k \right)_{k=0,...,T} $, as in (\ref{per_consumption}),
we denote
$
\widehat{c}_k :=  c(k, \beta , h) = c_k - \sum_{l=0}^{k-1} \beta^{(k)}_l c_l  - h_k ,
$
for all $k=0,...,T$.
\newline
\newline
\textbf{The case $k=T$.} Let $ \psi_{T} : L^{2} ( \mathcal{G}_T ) \to
L^{2} ( \mathcal{G}_T ) $ be a function defined by
\begin{equation}
\psi_{T} ( X ) := \epsilon_T + X.
\label{tec0}
\end{equation}
Observe that relation (\ref{first2}), for $k=T$, implies that $c^{*}_T  =
\psi_T ( W^{*}_T )$. Furthermore, $d \psi_{T} = 1 $. Next, consider the function
\begin{equation}
f_{T} ( c_0  , ... , c_{T-1} , W_{T} ) = P_{ \mathcal{L} }^{T}
\left[ u'_T \left(  \psi_{T}( W_T ) - \sum_{l=0}^{T-1} \beta^{(T)}_l c_l  - h_T \right) \right]
\end{equation}
\[
- \frac{ \widetilde{M}_T }{ \widetilde{M}_{T-1} }  u'_{T-1} \left( \widehat{c}_{T-1}  \right),
\]
defined on the set $ B'_T \subseteq L^{2} ( \mathcal{G}_0 ) \times
... \times L^2 ( \mathcal{G}_{T-1} ) \times \mathcal{L}_T $
consisting of all tuples $( c_0  , ... , c_{T-1} , W_{T} )$ such
that $  \psi_{T}( W_T ) - \sum_{l=0}^{T-1} \beta^{(T)}_l c_l  -
h_T \geq 0 $ and $  \widehat{c}_{T-1} \geq 0$, $P-$a.s.
Note that Lemma \ref{reformulation} implies that $ f_{T} ( c^{*}_0 , ... , c^{*}_{T-1} , W^{*}_{T} 
) =  0 $. Furthermore, the differential of $ f_{T}$ with respect
to $W_{T}$, at some $ (c_0 , ... , c_{T-1}, W_T ) \in int \left(
B_{T} \right)$, $\frac{ \partial f_T }{ \partial W_T } :=
\frac{ \partial f_T }{ \partial W_T }_{ (c_0 , ... , c_{T-1}, W_T )
}$, applied on some vector $V \in \mathcal{L}_T$, $V \neq 0$, is given by
\[
\frac{ \partial f_T }{ \partial W_T } V = P^{T}_{ \mathcal{L} } \left[ u''_{T} \left( \psi_{T} \left(
W_T \right)  - \sum_{l=0}^{T-1} \beta^{(T)}_l c_l  - h_T  \right)
V \right].
\]
Note that part (iii) of Lemma \ref{teclma1} yields
\[
E \left[  V \frac{ \partial f_T }{ \partial W_k }_{ (c_0 , ... , c_{T-1}, W_T ) } V \big| \mathcal{G}_{T-1} \right] =
\]
\[
E \left[ u''_{T} \left( \psi_{T} \left( W_T \right)  - \sum_{l=0}^{T-1} \beta^{(T)}_l c_l  - h_T  \right) V^2  \big| \mathcal{G}_{T-1} \right] < 0,
\]
where the last inequality follows by the concavity of $u_T.$ This in particular implies that the operator
\[
\frac{ \partial f_T
}{ \partial W_T }_{ (c^{*}_0 , ... , c^{*}_{T-1} , W^{*}_T  ) } : \mathcal{L}_T \to \mathcal{L}_T ,
\]
which is given by
\[ \frac{ \partial f_T }{
\partial W_T }_{ (c^{*}_0 , ... , c^{*}_{T-1} , W^{*}_T  ) } V := P^{\mathcal{L}}_{T} \left[ u''_{T}
\left( \widehat{c}^{*}_{T}   \right) V \right]
\]
is non-degenerate. The implicit function theorem implies that
there exists an open neighborhood $D_{T-1}$ of $( c^{*}_0 , ... , c^{*}_{T-1} )$ (in the $L^{2} ( \mathcal{G}_0
) \times ... \times L^2 ( \mathcal{G}_{T-1} )-$topology) and a
unique $C^1-$function $ \eta_{T} : D_{T-1} \to \mathcal{L}_T$ such that
\begin{equation}
f_{T} (  c_0 , ... , c_{T-1} , \eta_{T} ( c_0, ... , c_{T-1} )  ) = 0 ,
\label{tec12}
\end{equation}
for all $ ( c_0 , ... , c_{T-1} ) \in D_{T-1}$. We will compute now certain differentials
that will be crucial for proving the statement in the phase: $k=T-1$. A differentiation of (\ref{tec12})
with respect to $c_{T-1}$ and $c_{k}$, $k=0,...,T-2$, combined with item (ii) of Lemma \ref{teclma1},
implies the validity of the following identities,
\begin{equation}
P_{ \mathcal{L} }^{T} \left[ u''_T \left(  \psi_{T} ( \eta_T ( c_0
, ... , c_{T-1} ) )  - \sum_{l=0}^{T-1} \beta^{(T)}_l c_l  - h_T
\right) \left( \frac{ \partial \eta_T }{ \partial c_{T-1} } V  -
\beta^{(T)}_{T-1} V  \right) \right]
\label{tec1}
\end{equation}
\[
 = \frac{ \widetilde{M}_T }{ \widetilde{M}_{T-1} }  u''_{T-1}
\left( \widehat{c}_{ T-1}  \right) V ,
\]
for all random variables $V \in L^2 (G_{T-1}), $ and
\begin{equation}
P_{ \mathcal{L} }^{T} \left[ u''_T \left(  \psi_{T} ( \eta_T ( c_0
, ... , c_{T-1} ) )  - \sum_{l=0}^{T-1} \beta^{(T)}_l c_l  - h_T
\right) \left( \frac{1}{W} \frac{ \partial \eta_T }{ \partial c_{k} } W  - \beta^{(T)}_k \right) \right]
\label{tec2}
\end{equation}
\[
 = - \beta^{(T-1)}_{ k } \frac{ \widetilde{M}_T }{ \widetilde{M}_{T-1}
}  u''_{T-1} \left( \widehat{c}_{T-1} \right) ,
\]
for all non-vanishing random variables $W \in L^2 (G_{ k }), $
$k=0,...,T-2$, where $\frac{ \partial \eta_T }{ \partial c_{T-1} } := \frac{ \partial \eta_T }{ \partial c_{T-1} }_{ (
c_0 , ... , c_{T-1} ) } $ and $ \frac{ \partial \eta_T }{ \partial c_{k} } := \frac{ \partial \eta_T }{ \partial c_{k} }_{ (
c_0 , ... , c_{T-1} ) } $. Therefore, by multiplying equations (\ref{tec1})
and (\ref{tec2}) by $\beta^{(T-1)}_{k}$ and $V$ respectively, applying property (ii) in Lemma \ref{teclma1},
and summing up these equations, we obtain
\[
P_{ \mathcal{L} }^{T} \bigg[ u''_T \left(  \psi_{T} ( \eta_T ( c_0
, ... , c_{T-1} ) )  - \sum_{l=0}^{T-1} \beta^{(T)}_l c_l  - h_T
\right) 
\]
\[
 \times \left(  \beta^{(T-1)}_k  \frac{ \partial \eta_T
}{
\partial c_{T-1} } V + \frac{ V }{W} \frac{ \partial \eta_T }{ \partial c_{k} }W
\right)  \bigg] =
\]
\[
P_{ \mathcal{L} }^{T} \left[ u''_T \left(  \psi_{T} ( \eta_T ( c_0
, ... , c_{T-1} ) )  - \sum_{l=0}^{T-1} \beta^{(T)}_l c_l  - h_T
\right) \left( \beta^{ ( T-1 ) }_k \beta^{(T)}_{T-1} +
\beta^{(T)}_k  \right) V \right] ,
\]
and thus by part (iv) of Lemma \ref{teclma1}, it follows that
\begin{equation}
 \frac{ \partial \eta_T }{
\partial c_{T-1} } V   =  \frac{1}{ \beta^{(T-1)}_k } \left( \beta^{ ( T-1 ) }_k \beta^{(T)}_{T-1} + \beta^{(T)}_k  - \frac{ 1}{W} \frac{ \partial \eta_T }{ \partial c_{k} } W \right) V.
\label{tec3}
\end{equation}
By employing similar arguments, one can check that equation (\ref{tec2}) implies that
\begin{equation}
\frac{ \partial \eta_T }{ \partial c_k } W = \left( \beta^{(T)}_{k} +  \frac{ \beta^{(T -1 )}_{k} }{ \beta^{(T-1)}_{m } }
\left( \frac{1}{Y} \frac{ \partial \eta_T }{ \partial c_m } Y - \beta^{(T)}_{m}   \right)    \right) W ,
\label{tec4}
\end{equation}
for all $k , m = 0 , ... , T-2, $ $W \in L^2 \left(  \mathcal{G}_{k-1} \right) $
and all non-vanishing random variables $ Y \in L^2 \left(  \mathcal{G}_{m-1} \right) .$ In particular
we deduce that $  \frac{ \partial \eta_T }{ \partial c_{T-1} } $, and $  \frac{ \partial \eta_T }{ \partial c_{k} } $, for $k=0,...,T-2,$
are multiplication operators (and thus can be treated as random variables). That is, we can rewrite (\ref{tec3}) as
\begin{equation}
 \frac{ \partial \eta_T }{
\partial c_{T-1} }    =  \frac{1}{ \beta^{(T-1)}_k } \left( \beta^{ ( T-1 ) }_k \beta^{(T)}_{T-1} + \beta^{(T)}_k  - \frac{ \partial \eta_T }{ \partial c_{k} }
\right).
\label{tec13}
\end{equation}
\newline
\newline
\textbf{The case $k=T-1$.} We set the function
\[
g_{T-1} : D_{T-1} \times \mathcal{L}_{T-1} \to L^{2} \left(
\mathcal{G}_{T-1} \right) ,
\]
\[
g_{T-1} (c_0, ... , c_{T-1}, W_{T-1})  = c_{T-1} + E \left[
\frac{M_{T}}{M_{T-1}} \eta_T ( c_0 , ... , c_{T-1} ) |
\mathcal{G}_{T-1} \right]
\]
\[
- W_{T-1} - \epsilon_{T-1},
\]
and note that by (\ref{first2}), $g_{T-1} (c^{*}_0 , ... , c^{*}_{T-1} , W^{*}_{T-1} ) = 0 $. Let
\[
 \frac{ \partial g_{T-1} }{ \partial c_{T-1} } :=   \frac{ \partial g_{T-1} }{ \partial c_{T-1} }_{ (c_0, ... , c_{T-1}, W_{T-1}) } : L^{2} \left( \mathcal{G}_{T-1} \right) \to L^{2} \left( \mathcal{G}_{T-1}
  \right),
\]
be the derivative, which is given by
\[
 \frac{ \partial g_{T-1} }{ \partial c_{T-1} } V =  V + E \left[ \frac{M_T}{ M_{T-1} }  \frac{ \partial \eta_{T} }{ \partial c_{T-1} } V \big| \mathcal{G}_{T-1}
 \right] ,
\]
for all $ V \in L^{2} ( \mathcal{G}_{T-1} )  $. Observe that
\[
E \left[ V \frac{ \partial g_{T-1} }{ \partial
c_{T-1} } V | \mathcal{G}_{T-1} \right] =
V^2 +  E \left[ \frac{ M_T }{ M_{T-1} } V \frac{ \partial \eta_T }{ \partial c_{T-1} } V     | \mathcal{G}_{T-1} \right] .
\]
Let us show that $ E \left[ \frac{ M_T }{ M_{T-1} } V \frac{ \partial \eta_T }{ \partial c_{T-1} } V     | \mathcal{G}_{T-1} \right] > 0 $, for $V \neq 0.$
By (\ref{tec13}), we have
\[
E \left[ \frac{ M_T }{ M_{T-1} } V \frac{ \partial \eta_T }{ \partial c_{T-1} } V     | \mathcal{G}_{T-1} \right] =
\]
\[
\beta^{(T)}_{T-1} E \left[ \frac{ M_T }{ M_{T-1} }  | \mathcal{G}_{T-1} \right] V^2 + \frac{1}{ \beta^{(T-1)}_{k} }
E \left[ \frac{M_T}{ M_{T-1} } \left( \beta^{(T)}_k -  \frac{ \partial \eta_T }{ \partial c_k }  \right) | \mathcal{G}_{T-1}   \right] V^2.
\]
Recall that $ E \left[ \frac{ M_T }{ M_{T-1} }  | \mathcal{G}_{T-1} \right] $ denotes the interest rate at period $T-1$, and thus it is in particular a
positive quantity. Thereby, it suffices to show that
\[
E \left[ \frac{M_T}{ M_{T-1} } \left( \beta^{(T)}_k -  \frac{ \partial \eta_T }{ \partial c_k }   \right) | \mathcal{G}_{T-1}   \right] > 0.
\]
In virtue of (\ref{tec2}), this is equivalent to showing that
\begin{equation}
\frac{1}{ \beta^{(T-1)}_k u''_{T-1} \left(  \widehat{c}_{T-1}  \right) } E \bigg[  \frac{ M_T }{ \widetilde{M}_{T} } \frac{ M_{T-1} }{ \widetilde{M}_{T-1} } \left( \beta^{(T)}_{k} - \frac{ \partial \eta_T   }{ \partial c_k   }   \right) 
\label{tec5}
\end{equation}
\[
\times P^{T}_{ \mathcal{L}}  \bigg[ u''_{T} \left(  \psi_{T} ( \eta_T ( c_0
, ... , c_{T-1} ) )  - \sum_{l=0}^{T-1} \beta^{(T)}_l c_l  - h_T
\right) 
\]
\[
\times \left( \beta^{(T)}_{k} -  \frac{ \partial \eta_T   }{ \partial c_k   }   \right)   \bigg]  \big| \mathcal{ G }_{ T -1 }  \bigg] > 0.
\]
To this end, let us first prove it for the deterministic interest rate case, and then for idiosyncratically incomplete markets.
Note that in the former case, the quantities $ \frac{ M_k }{ \widetilde{M}_K } ,$ $k=1,...,T$ are positive numbers, and hence,
since $  \beta^{(T)}_{k} - \frac{1}{W} \frac{ \partial \eta_T   }{ \partial c_k   } W \in \mathcal{L}_{T} $,
part (iii) of Lemma \ref{teclma1}, implies that (\ref{tec5}) is equivalent to
\[
E \left[  u''_{T} \left(  \psi_{T} ( \eta_T ( c_0
, ... , c_{T-1} ) )  - \sum_{l=0}^{T-1} \beta^{(T)}_l c_l  - h_T
\right)  \left( \beta^{(T)}_{k} - \frac{ \partial \eta_T   }{ \partial c_k   }  \right)^2   \big| \mathcal{G}_{T-1} \right] < 0,
\]
which holds true since $u_T$ is a concave function. For idiosyncratically incomplete markets, notice first that the ratio $\frac{M_k}{\widetilde{M}_k}$ is positive,
for all $k=1,...,T$. Next, since $ P^{ \mathcal{L} }_{k} \left[ \cdot  \right] = E \left[ \cdot \big| \sigma \left( \mathcal{G}_{k-1} , \mathcal{F}_k \right) \right] $, it follows that $P^{ \mathcal{L} }_{k} \left[ X Y \right] = X P^{ \mathcal{L} }_{k} \left[ Y \right] $ holds, for all
$Y \in L^{2} \left( \mathcal{G}_T  \right) $ and $X \in \mathcal{L}_k = L^{2} \left( \sigma \left( \mathcal{G}_{k-1} , \mathcal{F}_k \right)  \right) .$ By exploiting these observations,
one easily checks that (\ref{tec5}) is equivalent to
\begin{equation}
 E \bigg[  u''_{T} \left(  \psi_{T} ( \eta_T ( c_0
, ... , c_{T-1} ) )  - \sum_{l=0}^{T-1} \beta^{(T)}_l c_l  - h_T
\right)
\label{tec21}
\end{equation}
\[
\times \left( \beta^{(T)}_{k} - \frac{ \partial \eta_T   }{ \partial c_k   }  \right)^2   \big| \sigma \left( \mathcal{G}_{k-1} , \mathcal{F}_k \right) \bigg] < 0,
\]
which is evidently satisfied. Therefore, it follows that the operator
\[
 \frac{ \partial g_{T-1} }{ \partial c_{T-1} }_{ ( c^{*}_0 , ..., c^{*}_{T-1} , W^{*}_{T-1}  )}  :
L^{2} \left( \mathcal{G}_{T-1} \right) \to L^{2} \left( \mathcal{G}_{T-1} \right),
\]
which is given by
\[
 \frac{ \partial g_{T-1} }{ \partial c_{T-1} }_{ ( c^{*}_0 , ..., c^{*}_{T-1} , W^{*}_{T-1}  )}  V =  
\]
\[
V + E \left[ \frac{M_T}{ M_{T-1} }  \frac{ \partial \eta_{T} }{ \partial c_{T-1} }_{ ( c^{*}_0 , ..., c^{*}_{T-2} , W^{*}_{T-1}) }  V \big| \mathcal{G}_{T-1} \right]
\]
is non-degenerate. The implicit function theorem implies that
there exists an open neighborhood $B_{T-1}$ of $ ( c^{*}_0 , ..., c^{*}_{T −2} , W^{*}_{T-1} )$
(in the $L^2 ( \mathcal{G}_0 ) \times ... \times L^2 (
\mathcal{G}_{T-2}) \times \mathcal{L}_{T-1}-$topology) and a
unique $C^1-$function $ \psi_{T-1} : B_{T-1} \to L^{2} (
\mathcal{G}_{T-2} ) $ such that
\[
 g_{ T -1} ( c_0  , ..., c_{T - 2 } , \psi_{T-1} ( c_0  , ..., c_{T -2} , W_{T-1} )  , W_{T-1}) =
\Psi_{T-1} ( c_0 , ... , c_{T-2} , W_{T-1} ) +
\]
\[
 E \left[  \frac{ M_T }{ M_{T-1} } \eta_{T} ( c_0 , ... , c_{T-2} ,
\psi_{T-1} ( c_0 , ... , c_{T-2} , W_{T-1} ) )  \big| \mathcal{G}_{T-1} \right]
 - W_{T-1} - \epsilon_{T-1} = 0 ,
\]
for all $( c_0  , ..., c_{T −2} , W_{T-1} ) \in B_{T-1}.$ By differentiating the above equation with respect to
$W_{T-1}$ and $c_k$, $k=0,...,T-2$, and using the fact that both $  \frac{ \partial \eta_{T} }{
\partial c_{T-1} }$ and $ \frac{ \partial \eta_{T} }{
\partial c_{k} }$ are multiplicative operators, we obtain that both $ \frac{ \partial \psi_{T-1} }{ \partial W_{T-1} } :=
\frac{ \partial \psi_{T-1} }{ \partial W_{T-1} }_{( c_0  , ..., c_{T - 2} , W_{T-1} )} $ and $ \frac{ \partial \psi_{T-1} }{ \partial c_{k} } :=
\frac{ \partial \psi_{T-1} }{ \partial c_{k} }_{( c_0  , ..., c_{T -2} , W_{T-1} )}  $ are multiplicative operators given explicitly by

\begin{equation}
\frac{ \partial \psi_{T-1} }{ \partial W_{T-1} } =
\frac{ 1 }{ 1 + E \left[ \frac{M_T}{M_{T-1} }  \frac{ \partial \eta_{T} }{
\partial c_{T-1} }_{( c_0  , ..., c_{T - 2} ,   \psi_{T-1} ( c_0  , ..., c_{T - 2} , W_{T-1} ) , W_{T-1} )} | \mathcal{G}_{T-1} \right] },
\label{tec9}
\end{equation}
and
\begin{equation}
\frac{ \partial \psi_{T-1} }{ \partial c_{k} } = \frac{ - E \left[ \frac{M_T}{M_{T-1} }  \frac{ \partial \eta_{T} }{ \partial c_{ k } }_{( c_0  , ..., c_{T - 2} ,   \psi_{T-1} ( c_0  , ..., c_{T - 2} , W_{T-1} ) , W_{T-1} )}
| \mathcal{G}_{T-1} \right] }{ 1 + E \left[ \frac{M_T}{M_{T-1} }  \frac{ \partial \eta_{T} }{
\partial c_{T-1} }_{( c_0  , ..., c_{T - 2} ,   \psi_{T-1} ( c_0  , ..., c_{T - 2} , W_{T-1} ) , W_{T-1} )} | \mathcal{G}_{T-1} \right] }.
\label{tec15}
\end{equation}
Unless unavoidable, we will henceforth omit the indexes in the derivatives. Next, we define the function,
\[
f_{T-1} ( c_0 , ... , c_{T-2}, W_{T-1} ) = - \frac{ \widetilde{M}_{T-1} }{ \widetilde{M}_{T-2} } u'_{T-2} \left( \widehat{c}_{T-2} \right)
\]
\[
+ P^{T-1}_{ \mathcal{L}} \left[ u'_{T-1} \left( \psi_{T-1} ( c_0 , ... , c_{T-2}, W_{T-1} ) - \sum_{ k=0 }^{ T-2} \beta^{(T-1)}_k c_k - h_{T-1}   \right)  \right]  ,
\]
on the set $ B_{T-1}$, which consists of all tuples $( c_0 , ... ,
c_{T-2}, W_{T-1} ) \in L^{2} \left( \mathcal{G}_0 \right) \times
... \times L^{2} \left( \mathcal{G}_{ T-2 } \right) \times
\mathcal{L}_{T-1} $ such that $ \psi_{T-1} ( c_0 , ... , c_{T-2},
W_{T-1} ) - \sum_{ k=0 }^{ T-2} \beta^{(T-1)}_k c_k - h_{T-1} \geq
0$ and $ \widehat{c}_{T-2} \geq $ 0. Observe that
\[
\frac{ \partial f_{T-1} }{ \partial W_{T-1} } V =
\]
\[
P^{T-1}_{ \mathcal{L}} \left[ u''_{T-1} \left( \psi_{T-1} ( c_0 , ... , c_{T-2}, W_{T-1} ) - \sum_{ k=0 }^{ T-2} \beta^{(T-1)}_k c_k - h_{T-1}   \right)
\frac{ \partial \psi_{T-1} }{ \partial W_{T-1} } V  \right],
\]
for all $V \in \mathcal{L}_{T-1} $. By item (iii) in Lemma \ref{teclma1}, it follows that
\[
E \left[ V \frac{ \partial f_{T-1} }{ \partial W_{T-1} }V   \big| \mathcal{G}_{T-2} \right] =
\]
\[
E \bigg[  u''_{T-1} \left( \psi_{T-1} ( c_0 , ... , c_{T-2}, W_{T-1} ) - \sum_{ k=0 }^{ T-2} \beta^{(T-1)}_k c_k - h_{T-1}   \right) 
\]
\[
\times \frac{ \partial \psi_{T-1} }{ \partial W_{T-1} }  V^2    \big| \mathcal{G}_{T-2} \bigg] < 0,
\]
where the last inequality follows by the concavity of $u_{T-1}$ and the fact that
$ \frac{ \partial \psi_{T-1} }{ \partial W_{T-1} } > 0,$ which holds due to (\ref{tec9}).
Therefore, we conclude that the differential
\[
\frac{ \partial f_{T-1} }{ \partial W_{T-1} }_{ ( c^{*}_0 , ... , c^{*}_{T-2} , W^{*}_{T-1}  ) } : \mathcal{L}_{T-1} \to \mathcal{L}_{T-1},
\]
which is given by
\[
\frac{ \partial f_{T-1} }{ \partial W_{T-1} }_{ ( c^{*}_0  , ... , c^{*}_{T-2}  , W^{*}_{T-1}  ) }  V =
\]
\[
P^{T-1}_{ \mathcal{L}} \left[ u''_{T-1} \left( \widehat{c}^{*}_{T-1}   \right)
\frac{ \partial \psi_{T-1} }{ \partial W_{T-1} }_{ ( c^{*}_0  , ... , c^{*}_{T-2} , W^{*}_{T-1}   ) } V  \right]
\]
is non-degenerate. The implicit function theorem implies the existence of an open neighborhood $D_{T-2}$ of
$(c^{*}_0, ... , c^{*}_{T-2}  )$ (in the $L^{2}( \mathcal{G}_0 ) \times ... \times L^{2} ( \mathcal{G}_{T-2} )-$topology)
and a unique $C^1-$function $ \eta_{T-1}( c_0 , ... , c_{T-2} ): D_{T-2} \to \mathcal{L}_{T-1}$, such that
\[
f_{T-1} ( c_0 , ... , c_{T-2}, \eta_{T-1}( c_0 , ... , c_{T-2} ) ) = 0,
\]
for all $ (c_0 , ... , c_{T-2}) \in D_{T-2} .$ We shall compute
now some differentials and establish related identities that we will find useful in the subsequence phases. By
differentiating the above equation with respect to $c_{T-2}$ and
then $c_k$, $k=0,...,T - 3,$ one checks that the following is
satisfied:
\begin{equation}
P^{T-1}_{\mathcal{L}} \bigg[ u''_{T-1} \left( \psi_{T-1} ( c_0 ,
... , c_{T-2}, W_{T-1} ) - \sum_{ k=0 }^{ T-2} \beta^{(T-1)}_k c_k
- h_{T-1} \right) 
\label{111}
\end{equation}
\[
\times
\left( \frac{ \partial \psi_{T-1} }{ \partial c_{T-2} } \cdot V +
\frac{ \partial \psi_{T-1} }{ \partial W_{T-1} }  \cdot \frac{
\partial \eta_{T-1} }{ \partial c_{T-2} }V  - \beta^{(T-1)}_{T-2}
\cdot V \right)   \bigg] =  \frac{ \widetilde{ M}_{T-1}  }{
\widetilde{ M}_{T-2} }  u'_{T-2} \left( \widehat{c}_{T-2} \right)
\cdot V ,
\]
for all $V \in L^{2} \left( \mathcal{G}_{ T-2 } \right), $ and
\begin{equation}
P^{T-1}_{\mathcal{L}} \bigg[ u''_{T-1} \left( \psi_{T-1} ( c_0 , ... , c_{T-2}, W_{T-1} ) - \sum_{ k=0 }^{ T-2} \beta^{(T-1)}_k c_k - h_{T-1} \right) 
\label{tec8}
\end{equation}
\[
 \times \left( \frac{ \partial \psi_{T-1} }{ \partial c_{k} }
+ \frac{ \partial \psi_{T-1} }{ \partial W_{T-1} }  \cdot
\frac{1}{W} \frac{ \partial \eta_{T-1} }{ \partial c_{k} } W -
\beta^{(T-1)}_{k}  \right)   \bigg] = - \beta^{(T-2)}_{k} \frac{
\widetilde{ M}_{T-1}  }{ \widetilde{ M}_{T-2} }  u'_{T-2} \left(
\widehat{c}_{T-2} \right),
\]
for all non-vanishing vectors $W \in L^{2} \left( \mathcal{G}_k \right) $, $k=0,...,T-3$. Similarly as in (\ref{tec3}) and (\ref{tec4}),
by part (iv) of Lemma \ref{teclma1}, we obtain that
\[
\beta^{(T-2)}_k \left( \frac{ \partial \psi_{T-1} }{ \partial c_{T-2} } \cdot V
+ \frac{ \partial \psi_{T-1} }{ \partial W_{T-1} }  \cdot \frac{ \partial \eta_{T-1} }{ \partial c_{T-2} } V - \beta^{(T-1)}_{T-2} \cdot V  \right)
\]
\[
+ \left( \frac{ \partial \psi_{T-1} }{ \partial c_{k} }
+ \frac{1}{W} \frac{ \partial \psi_{T-1} }{ \partial W_{T-1} }  \cdot \frac{ \partial \eta_{T-1} }{ \partial c_{k} } W - \beta^{(T-1)}_{k}  \right) \cdot V
=0,
\]
or equivalently
\[
\frac{ \partial \eta_{T-1} }{ \partial c_{T-2} } = \bigg( \beta^{(T-1)}_{T-2} - \frac{ \partial \psi_{T-1} }{ \partial c_{T-2} }
- \frac{1 }{ \beta^{(T-2)}_k }  \bigg(  \frac{ \partial \psi_{T-1} }{ \partial c_{k} }
+
\]
\[
\frac{1}{W} \frac{ \partial \psi_{T-1} }{ \partial W_{T-1} }  \cdot \frac{ \partial \eta_{T-1} }{ \partial c_{k} }W - \beta^{(T-1)}_{k}
  \bigg)  \bigg) \frac{1}{ \frac{ \partial \psi_{T-1} }{ \partial W_{T-1} } }.
\]
We turn now to a simplification of some terms in the brackets above,
\begin{equation}
 \beta^{(T-1)}_{T-2} - \frac{ \partial \psi_{T-1} }{ \partial c_{T-2} } =
\frac{ \beta^{(T-1)}_{T-2} + E \left[ \frac{ M_T }{ M_{T-1}} \left( \beta^{(T-1)}_{T-2} \frac{ \partial \eta_T }{ \partial c_{T-1} } + \frac{ \partial \eta_T }{ \partial c_{T-2} }   \right)     \big| \mathcal{ G }_{T-1} \right]   }{ 1 + E \left[ \frac{ M_T }{ M_{T-1} } \frac{  \partial \eta_{T} }{ \partial c_{T-1} }    \big| \mathcal{ G }_{T-1}  \right] } =
\label{tec6}
\end{equation}
\[
\frac{ \beta^{(T-1)}_{T-2} +   \left(  \beta^{(T-1)}_{T-2} \beta^{(T)}_{T-1} + \beta^{T}_{(T-2)}  \right)  E \left[ \frac{ M_T }{ M_{T-1}} \big| \mathcal{G}_{T-1}  \right]   }{ 1 + E \left[ \frac{ M_T }{ M_{T-1} } \frac{  \partial \eta_{T} }{ \partial c_{T-1} }    \big| \mathcal{ G }_{T-1}  \right] },
\]
where the first equation follows by (\ref{tec15}) and the second one is due to (\ref{tec4}).
Next, by (\ref{tec9}), (\ref{tec15}) and (\ref{tec4}), we have
\begin{equation}
X := \frac{1}{ \frac{ \partial \psi_{T-1} }{ \partial W_{T-1} } } \left(  \frac{ \partial \psi_{T-1} }{ \partial c_{k} }
- \beta^{(T-1)}_{k} + \frac{1}{W} \frac{ \partial \psi_{T-1} }{ \partial W_{T-1} }  \cdot \frac{ \partial \eta_{T-1} }{ \partial c_{k} }W  \right)
=
\label{tec7}
\end{equation}
\[
 - E \left[ \frac{ M_T }{ M_{T-1} } \left(  \frac{ \partial \eta_T }{ \partial c_k } + \beta^{(T-1)}_k   \frac{ \partial \eta_T }{ \partial c_{T-1} }   \right)  \big| \mathcal{G}_{T-1} \right]  +  \frac{1}{W} \frac{ \partial \eta_{T-1} }{ \partial c_{k} } W =
\]
\[
 - \left(  \beta^{ ( T-1 ) }_{k} \beta^{(T)}_{T-1} + \beta^{(T)}_k  \right) E \left[ \frac{ M_T }{ M_{T-1} }   \big| \mathcal{G}_{T-1} \right]  +  \frac{1}{W} \frac{ \partial \eta_{T-1} }{ \partial c_{k} } W.
\]
Note that the above computation implies in particular that $ X \in
\mathcal{L}_{T-1}$, in both cases of a deterministic interest rate
and idiosyncratically incomplete markets. Indeed, in both cases $\frac{1}{W}
\frac{ \partial \eta_{T-1} }{ \partial c_{k} } W \in
\mathcal{L}_{T-1}$. In the former case, $ E \left[ \frac{ M_T
}{ M_{T-1} }   \big| \mathcal{G}_{T-1} \right]$ is a real number
indicating the interest rate, and in the latter case, we have $ E
\left[ \frac{ M_T }{ M_{T-1} }   \big| \mathcal{G}_{T-1} \right] =
E \left[ \frac{ M_T }{ M_{T-1} }   \big| \mathcal{F}_{T-1} \right]
\in \mathcal{L}_{T-1} := L^{2} \left( \sigma \left(
\mathcal{G}_{T-2} ,  \mathcal{F}_{T-1} \right) \right) $, by item
(iii) in Definition \ref{def1}, and Lemma \ref{lem1}. Therefore,
by (\ref{tec6}) and (\ref{tec7}), we have
\[
\frac{ \partial \eta_{T-1}  }{ \partial c_{T-2} } = - \frac{1}{ \beta^{(T-2)}_k }  X +  \beta^{(T-1)}_{T-2} +   \left(  \beta^{(T-1)}_{T-2} \beta^{(T)}_{T-1} + \beta^{(T)}_{T-2}  \right)  E \left[ \frac{ M_T }{ M_{T-1}} \big| \mathcal{G}_{T-1}  \right].
\]
Let us show that $ E \left[ - X  \frac{ M_{T-1} }{ M_{T-2} } \big| \mathcal{G}_{T-2}  \right] > 0. $ This will imply that
\begin{equation}
E \left[ \frac{ \partial \eta_{T-1}  }{ \partial c_{T-2} }  \frac{ M_{T-1} }{ M_{T-2} }  \big| \mathcal{G}_{T-2} \right]  >
\beta^{(T-1)}_{T-2} E \left[  \frac{ M_{T-1} }{ M_{T-2} }  \big| \mathcal{G}_{T-2} \right] +
\label{tec20}
\end{equation}
\[
\left(   \beta^{(T-1)}_{T-2} \beta^{(T)}_{T-1} + \beta^{ (T)}_{T-2}  \right) E \left[  \frac{ M_{T} }{ M_{T-2} }  \big| \mathcal{G}_{T-2} \right].
\]
One checks by using relation (\ref{tec8}) that the claim is equivalent to
\begin{equation}
E \bigg[
X  \frac{ M_{T-1} }{ M_{T-2} }
\frac{ \widetilde{M}_{T-2} }{ \widetilde{M}_{T-1} } \frac{1}{ u''_{T-2} ( \widehat{c}_{T-2}  ) }
 P^{\mathcal{L}}_{T-1} \left[ u''_{T-1} \left( Y \right)   \frac{\partial \psi_{T-1} }{ \partial W_{T-1} } X    \right] \bigg| \mathcal{G}_{T-2}   \bigg] > 0,
\label{tec}
\end{equation}
where,
\[
Y := \psi_{T-1} ( c_0 , ... , c_{T-2}, W_{T-1} ) - \sum_{ k=0 }^{ T-2} \beta^{(T-1)}_k c_k - h_{T-1}  .
\]
For a market with a deterministic interest-rate, recall that $
\frac{ \widetilde{M}_k }{M_k} ,$ is a positive number, for each
$k$. Therefore, since $X \in \mathcal{L}_{T-1} $, part (iii) of
Lemma \ref{teclma1} implies that (\ref{tec8}) is equivalent to
\[
E  \left[  \frac{ \partial \psi_{T-1} }{ \partial W_{T-1} } X^2
 u''_{T-1} \left( Y  \right)
\big| \mathcal{G}_{T-2} \right] < 0,
\]
which holds true due to the assumption that $ u_{T-1} $ is a
concave function and that $ \frac{ \partial \psi_{T-1} }{ \partial
W_{T-1} } \geq 0,$ by (\ref{tec9}). For idiosyncratically incomplete markets,
since  $ \frac{ \widetilde{M}_k }{M_k} $ is a positive random
variable, for each $k$, and the operator $P^{T-1}_{\mathcal{L}}$
is a conditional expectation, we conclude (as in (\ref{tec21})) that
the validity of (\ref{tec}) is equivalent to verifying that
\[
E \left[ u''_{T-1} \left( Y \right)   \frac{\partial \psi_{T-1} }{ \partial W_{T-1} } X^2  \bigg|  \sigma \left(  \mathcal{G}_{T-2} ,  \mathcal{F}_{T-1} \right)
\right] < 0,
\]
which is satisfied due to the concavity of $u_{T-1}.$
\newline
\newline
\textbf{The General Case.} First, let us prove the statement for $k=T-2$
by exploiting the above results, and then briefly describe the general case, which is treated analogously.
Consider the function
\[
g_{T-2} : D_{T-2} \times \mathcal{L}_{T-2} \to L^{2} \left( \mathcal{G}_{T-2} \right),
\]
\[
g_{T-2}( c_0 , ... , c_{T-2} , W_{T-2} ) = c_{T-2} +
\]
\[
E \left[  \frac{M_{T-1} }{ M_{T-2} } \eta_{T-1} \left(  c_0 , ... , c_{T-2}
\right)   \big| \mathcal{G}_{T-2} \right] - W_{T-2} - \epsilon_{T-2} ,
\]
and note that (\ref{tec20}) implies that $ \frac{ \partial g_{T-2} }{
\partial c_{T-2} }_{ \left(  c^{*}_0 , ... , c^{*}_{T-2}
, W^{*}_{T-2}  \right) } : \mathcal{L}_{T-2}
\to \mathcal{L}_{T-2} $ is a non-degenerate linear operator, and thus
there exists a unique $C^1-$differentiable function $ \psi_{T-2}
:B_{T-2} \to L^{2} \left( \mathcal{G}_{T-2} \right) $, such that
\[
g_{T-2}( c_0 , ... , c_{T-3} , \psi_{T-2} \left( c_0 , ... , c_{T-3} , W_{T-2} \right)  , W_{T-2} ) = 0 .
\]
Observe that a differentiation of the above equation with respect to $W_{T-2}$ combined with (\ref{tec20}), yields

\begin{equation}
\frac{ \partial \psi_{T-2} }{ \partial W_{T-2} } =  \frac{1}{1 +
E \left[ \frac{ M_{ T -1 } }{ M_{T-2} } \frac{ \partial \eta_{T-1 }}{ \partial c_{T-2} }    \big| \mathcal{G}_{T-2} \right] }
\label{tec10}
\end{equation}
\[
\leq \frac{1}{1 +
\beta^{(T-1)}_{T-2} E \left[  \frac{ M_{T-1} }{ M_{T-2} }  \big| \mathcal{G}_{T-2} \right] +
\left(   \beta^{(T-1)}_{T-2} \beta^{(T)}_{T-1} + \beta^{(T)}_{T-2}  \right) E \left[  \frac{ M_{T} }{ M_{T-1} }  \big| \mathcal{G}_{T-2} \right] }.
\]
Finally, induction implies the existence of the following maps
\begin{equation}
f_k : B'_k \subseteq L^{2} \left(  \mathcal{G}_0 \right) \times ... \times L^2 \left( \mathcal{G}_{k-1} \right) \times \mathcal{L}_T  \to \mathcal{L}_T ,
\label{expl1}
\end{equation}
\[
f_k \left( c_0 , ... , c_{k-1} , W_k   \right) = P^{k}_{ \mathcal{L} } \left[ u'_k \left( \psi_k \left( c_0 , ... , c_{k-1} , W_k \right)
- \sum_{j=0}^{ k-1 } \beta^{(k)}_j c_j - h_k \right)  \right]
\]
\[
- \frac{ \widetilde{M}_k }{ \widetilde{M}_{k-1} } u'_{k-1} \left(  \widehat{c}_{k-1} \right) ,
\]
for $k=1,...,T-2,$ where $B'_k $ is the set of all tuples $ \left(
c_0 , ... , c_{k-1} , W_k   \right) \in B_k $ such that $ \psi_k
\left( c_0 , ... , c_{k-1} , W_k \right) - \sum_{j=0}^{ k-1 }
\beta^{(k)}_j c_j - h_k \geq 0 $ and $ \widehat{c}_{k-1} \geq 0;$
\begin{equation}
\eta_k : D_{k-1} \subseteq L^{2} \left( \mathcal{G}_0 \right) \times ... \times L^{2} \left( \mathcal{G}_{k-1} \right) \to \mathcal{L}_k
\label{expl2}
\end{equation}
such that
\[
f_k \left( c_0 , ... , c_{k-1} , \eta_k \left( c_0 , ... , c_{k-1} \right)   \right) = 0 ,
\]
for all $ \left( c_0 , ... , c_{k-1} \right) \in D_{k-1} $, $k=1,...,T-1$, where $D_{k-1}$ is some open neighborhood of $ \left( c^{*}_0 , ... , c^{*}_{k-1} \right)$ (in the $ L^{2} \left( \mathcal{G}_0 \right) \times ... \times L^{2} \left( \mathcal{G}_{k-1} \right)-$topology);
\begin{equation}
g_{k-1} : D_{k-1} \times \mathcal{L}_{k-1} \to L^{2} \left( \mathcal{G}_{k-1} \right)
\label{expl3}
\end{equation}
\[
g_{k-1} \left(c_0 , ... , c_{k-1} , W_{k-1}  \right)  = c_{k-1} + E \left[  \frac{M_k}{M_{k-1}}  \eta_k \left( c_0 , ... , c_{k-1} \right) \big| \mathcal{G}_{k-1} \right] - W_{k-1} - \epsilon_{k-1} ,
\]
and
\begin{equation}
\psi_{k-1} : B_{k-1}  \to L^{2} \left( \mathcal{G}_{k-1} \right)
\label{expl4}
\end{equation}
such that
\[
g_{k-1} \left( c_0 , ... , c_{k-2}, \psi_{k-1} \left( c_0 , ... , c_{k-2}, W_{k-1} \right)  , W_{k-1}  \right) = 0 ,
\]
for $k=2,...,T-1$, where $ B_{k-1}$ is some open neighborhood of $\left( c^{*}_0, ... , c^{*}_{k-2}, W^{*}_{k-1} \right) $
in the $L^{2} \left( \mathcal{G}_0 \right) \times ... \times L^{2} \left( \mathcal{G}_{k-2} \right) \times \mathcal{L}_{k-1}-$topology. In the last stage, we
set
\[
g_{0} ( c_0 , \epsilon_0 ) = c_0 + E \left[ M_1 \eta_1 \left( c_0 \right)  \right] - \epsilon_0,
\]
and 
\begin{equation}
\psi_0 : L^{2}_{+} \left(\mathcal{G}_{0} \right) \to L^{2}_{+} \left(\mathcal{G}_{0} \right) ,
\label{expl5}
\end{equation}
(where $ L^{2}_{+} \left(\mathcal{G}_{0} \right)  \cong {R}_{+}$), such that
\[
g_{0} ( c_0 , \psi_0 ( \epsilon_0 ) ) = 0 ,
\]
for all $ \epsilon_0 \in L^{2}_{+} \left(\mathcal{G}_{0} \right).$ This completes the proof. \qed
${}$
\newline
A further extension of the main result dealing with a
characterization of the response of the wealth to consumption, for
certain models of markets, is stated below.

\begin{Definition} An incomplete market is said to be of class $\mathcal{C}$, if
there exists an intermediate filtration $\left( \mathcal{H}_k
\right)_{ k=1,...,T}$ such that
\[
 G_{k-1} \subseteq \mathcal{H}_k
\subseteq \mathcal{G}_k,
\]
and $ P^{k}_{\mathcal{L}} \big[ \cdot \big] = E
\big[ \cdot  | \mathcal{H}_t \big],$ for all $ k=1, ... , T $.
\label{marketc}
\end{Definition}

\begin{Theorem} For an incomplete market of type $\mathcal{C}$ with a deterministic
interest rate, or for an idiosyncratically incomplete market, under the notations of Theorem
(\ref{main1}), we have
\[
\frac{ \partial \eta_{k+1} }{ \partial c_k } \geq
\sum_{j=k+1}^{ T }
\left( \sum_{i=k}^{j} \beta_{k}^{ i } \beta_{i}^{ i+1 } ... \beta_{j-1}^{j}   \right) E \left[ \frac{ M_j }{ M_{k+1} }  \big| \mathcal{G}_{k+1} \right],
\]
for all $k=0,...,T-1.$ \label{main2}
\end{Theorem}

As in the proof of Theorem \ref{main1}, our approach is
applicable in both models simultaneously. We will prove the above
statement for $k=T, T-1,$ whereas the rest can be easily completed
by induction.
\newline
\newline
\textbf{Proof of Theorem \ref{main2}} The key idea of the proof
is based on the algebraic identities developed in the proof of Theorem
\ref{main1} combined with the observation that the operators
$P^{k}_{\mathcal{L}},$ $k=1,...,T,$ are conditional expectations
in the setting of the current theorem. Observe that identity
(\ref{tec1}) accepts the form
\begin{equation}
\frac{ \partial \eta_T }{ \partial c_{T-1} } =
\label{tec31}
\end{equation}
\[
 = \beta^{(T)}_{T-1} + \frac{ \widetilde{M}_T }{ \widetilde{M}_{T-1} }
 \frac{u''_{T-1} \left( \widehat{c}_{ T-1}  \right)}{P_{ \mathcal{L} }^{T} \left[
u''_T \left(  \psi_{T} ( \eta_T ( c_0 , ... , c_{T-1} ) ) -
\sum_{l=0}^{T-1} \beta^{(T)}_l c_l  - h_T \right) \right]},
\]
proving the statement for $k=T$. Next, one can check by using
identity (\ref{tec6}), that $(\ref{111})$ can be transformed into
\begin{equation}
 \frac{ \partial \eta_{T-1} }{ \partial c_{T-2} }
= \frac{1}{ \frac{ \partial \psi_{T-1} }{ \partial W_{T-1} }
}\left( \beta^{(T-1)}_{T-2} - \frac{ \partial \psi_{T-1} }{
\partial c_{T-2} } \right)  +  \frac{ \widetilde{ M}_{T-1}  }{ \widetilde{
M}_{T-2} } 
\label{tec32}
\end{equation}
\[
\times 
\frac{ u'_{T-2} \left( \widehat{c}_{T-2} \right)}{
P^{T-1}_{\mathcal{L}} \bigg[
\frac{ \partial \psi_{T-1} }{ \partial W_{T-1} }
u''_{T-1} \left( \psi_{T-1} ( c_0 ,
... , c_{T-2}, W_{T-1} ) - \sum_{ k=0 }^{ T-2} \beta^{(T-1)}_k c_k
- h_{T-1} \right) \bigg] }.
\]
Now, note that identities (\ref{tec6}) and (\ref{tec9}) guarantee the validity of the
assertion for the case $k=T-1$. This completes the proof. $\qed$

\section{Concavity Results in Incomplete Markets}

In the previous section, \textit{arbitrary} time-inconsistent
utility functions served as generic preference functionals,
since the monotonicity feature of the consumption was valid for a large 
class of markets. As it is shown in the next two subsections,
the concavity property of the consumption is invalid for some
elementary and deterministic models of complete markets, when
general preferences are involved. Therefore, we will further
restrict ourselves only to some particular preferences
(time-consistent power utility functions), in order to ensure 
the concavity property for a rich variety of financial markets
(see Section 5.3).

\subsection{A Counterexample with Time-Inconsistent Preferences}

Consider a one-period complete market represented by the positive
SPD: $M_0=1$ and some $M_1>0.$ Assume that the investor is represented by
the utility functions $u_{0}(x) = \log x $, $u_{1}(x) = - x^{-1
} $ and $\rho=0. $The endowments of the agent are denoted by $\epsilon_0$ and
$\epsilon_1$; the habit coefficient is assumed to be
$\beta^{(1)}_{0} = 1$; exogenous habits are not incorporated,
i.e., $h_1=0$. We let $ c_0 ( \epsilon_0 )$ and $ c_1 ( \epsilon_0
)$ represent the optimal consumption stream. By Theorem
\ref{complete_gen_ut}, it follows that
\[
c_1 ( \epsilon_0 )  = c_0 ( \epsilon_0 ) + \sqrt{ c_0 ( \epsilon_0 )  M_1 }.
\]
The budget constraint $ c_0 ( \epsilon_0 ) + E [ M_1 c_1 (
\epsilon_0 ) ] =   \epsilon_0  + E [ M_1 \epsilon_1 ]$, implies
that
\[
 c_0 ( \epsilon_0 ) = \left( \frac{ \sqrt{ 4 a \epsilon_0 + 4ac + b^2} - b }{ 2a } \right)^2,
\]
where $  a = 1 + E[M_1],$ $ b = E [ M_1 \sqrt{ M_1 } ] , $
and $ c = E[M_1 \epsilon_1]. $ We have,
\[
0 < c'_0 ( \epsilon_0 ) = \frac{ 1}{ a } \frac{ \sqrt{ 4 a \epsilon_0 + 4ac + b^2} - b }{ \sqrt{ 4 a \epsilon_0 + 4ac + b^2} }  < 1 ,
\]
and
\[
c''_0 ( \epsilon_0 ) = 2b  \left(  4 a \epsilon_0 + 4ac + b^2 \right)^{ - 3 / 2 } > 0 .
\]
Thus $ c_0 (\epsilon_0 )$ is a \textit{convex} function.

\subsection{Why only Power Utility Functions?}
Consider a one-period complete market which consists only of a single riskless bond paying an interest rate
$r \in {R}_{+}.$ Assume that the individual is represented by the initial endowment
$ \epsilon_0 > 0 $ and some time-consistent utility function $ u  : {R}_{+} \to {R} ,$ which
satisfies the Inada conditions. We assume that no habits are involved. The corresponding utility maximization problem is:
\[
\sup_{ 0 \leq \pi_o \leq \epsilon_0  } u \big( \epsilon_0 - \pi_0 \big) + u \big( \pi_0
  r  \big).
\]
Let $\pi_0 (\epsilon_0) $ and $c_0 (\epsilon) := \epsilon_0 - \pi_0 (\epsilon_0) $ denote the optimal
portfolio and consumption respectively. The first order conditions imply that $\pi_0 \left(  \epsilon_0 \right) $ is determined
uniquely as the solution of the equation $ F \left( \epsilon_0   , \pi_0 ( \epsilon_0 )  \right) = 0 , $
where
\[
F \left( \epsilon_0   , \pi_0  \right) =  r  u' \left( \pi_0  r \right) - u' \left( \epsilon_0 - \pi_0  \right)  ,
\]
and hence, we have
\[
\pi_0 ( \epsilon_0 ) = \epsilon_0 - I ( r u' ( \pi_0 ( \epsilon_0 ) r)  ),
\]
where $ I(x) := (u' )^{-1} (x)$. The next statement demonstrates that \textit{only} power utility functions
imply the concavity of the consumption property.

\begin{Theorem}
The function $ c_0 ( \epsilon_0 )$ is concave for all $r \in
{R}_+$, if and only if $u(x) = c_1 + c_2   x^{ 1 - \gamma}
$, for arbitrary $ c_1, c_2 \in {R}$ and $ \gamma>0$.
Furthermore, in this case,
\[
c_0 ( \epsilon_0 ) = \frac{ r^{ 1 - 1 / \gamma } }{ 1 + r^{ 1 - 1 / \gamma } } \epsilon_0.
\]
Therefore, the only possible form of concavity here, is linearity.
\label{th1}
\end{Theorem}
\textbf{Proof of Theorem \ref{th1}.} Note that the concavity of $c_0( \epsilon_0)$ is equivalent
to the convexity of $ \pi_0 ( \epsilon_0 ) $. Observe that the function $\pi_0 ( \epsilon_0 ) $ is increasing, since
\[
\pi'_0 ( \epsilon_0 ) = \frac{ 1 }{  1 + r^{2} I' ( r  u' ( \pi_0 ( \epsilon_0 )  ) )  u'' ( \pi_0 ( \epsilon_0 ) r ) } > 0.
\]
Therefore, it follows that the function $\pi_0 ( \epsilon_0 )$ is convex for all $r>0,$ if and only if
the function $h(r,x) := I ( r u'( r x )) $ is concave with respect to the variable $x$, for all $r>0$.
Note that
\[
 \frac{ \partial h}{ \partial r} (r , x ) = I'(r  u' ( r  x) )  \left( u' ( r  x ) + r  u'' ( r  x )  x   \right).
\]
The identity $I(u'(x))=x,$ implies that $ I' ( u' ( x) ) = \frac{1}{ u''(x) }$, and thus
\[
 \frac{ \partial h}{ \partial r} (r , x )|_{r=1} = x + \frac{u'(x) }{ u''(x) }.
\]
On the other hand, since $h( 1 , x ) = 1,$ we have
\begin{equation}
\frac{ \partial h}{ \partial r} (r , x )|_{r=1} = \lim_{r \to 1} \frac{ h(r,x) - 1 }{ r- 1 }.
\label{concave_limit}
\end{equation}
The function $h(r,x) $ is a concave function of $x$, for all
$r>0$, if and only if the function $ \frac{ h(r,x) - h(1 , x) }{
r- 1 } $ is a concave function of $x$, for all $r > 1$, and a
convex function of $x$, for all $r< 1$. Therefore, it follows that
$ x + \frac{u'(x) }{ u''(x) }$ is a pointwise limit of concave and
convex functions, and thus we have $ \frac{u'(x) }{ u''(x) } = ax
+ b $, for some $ a , b \in R.$ This implies that $u'(x) = c  |a x
+ b |^{1 / a}, $ for some $c \in {R}$. The Inada condition
$u'(0) = \infty$ implies that $b = 0 $ and $a<0.$ Finally, we
conclude that $u(x) = c_1 + c_2   x^{ 1 - \gamma} $, for some $
c_1, c_2 \in R$ and $ \gamma = - 1 / a$, as required. \qed

\subsection{Concavity of The Optimal Consumption Stream}

We establish the concavity property of the optimal consumption stream
for an agent whose habit-forming preference are represented by a power utility,
in the setting of arbitrary idiosyncratically incomplete markets, and markets of type $\mathcal{C}$ 
with a deterministic interest rate (see Definition \ref{marketc}).

\begin{Theorem}
Given an idiosyncratically incomplete market, or a market of type $\mathcal{C}$ with a deterministic interest rate, and an agent represented by the utility functions
$u_k(x)=e^{- \rho k} \frac{ x^{1 - \gamma }}{ 1 - \gamma} $, $k=0,...,T,$ $ \rho \geq 0$, $\gamma \geq 0$, the optimal consumption is a concave function of wealth. That is, under the notations of Theorem (\ref{main1}), we have
\[
\frac{ \partial^2 \psi_k }{ \partial^2 W_k } \leq 0,
\]
for all $k=1,...,T-1,$ and
\[
\psi''_0 ( \epsilon_0 ) \leq 0.
\]
\label{main3}
\end{Theorem}

As in the proof of Theorem \ref{main1}, we will provide a proof for
the benchmark cases: $k=T-1, T-2$. A proof for a general period can be
carried out analogously. The proof applies to both models of markets.
\newline
\newline
\textbf{Proof of Theorem \ref{main3}.} A differentiation of equation (\ref{tec9}) with respect to $W_{T-1}$, yields
\[
\frac{ \partial^2 \psi_{T-1} }{ \partial^2 W_{T-1} } = - \frac{1}{  1 + \left( E \left[  \frac{ M_T }{ M_{T-1} }  \frac{ \partial \eta_T }{ \partial c_{T-1} }  \big|  \mathcal{G}_{T-1}  \right]     \right)^2  }
E \left[ \frac{ M_T }{ M_{T-1} }  \frac{ \partial^2 \eta_{T} }{ \partial^2 c_{T-1} } \frac{ \partial \psi_{T-1} }{ \partial W_{T-1} }   \bigg| \mathcal{G}_{T-1} \right] .
\]
By differentiating equation (\ref{tec31}) with respect to $c_{T-1}$, we get
\[
\frac{ \partial^2 \eta_{T} }{ \partial^2 c_{T-1} } = \frac{ \widetilde{M}_T }{ \widetilde{M}_{T-1} }
 \frac{u'''_{T-1} \left( X  \right) P_{ \mathcal{L} }^{T} \left[
u''_T \left(  Y  \right) \right] - u''_{T-1} ( X  ) P_{ \mathcal{L} }^{T} \left[
u''_T \left(  Y  \right) \right] \left(   \frac{ \partial \eta_T }{ \partial c_{T-1} } - \beta^{(T)}_{T-1}  \right)  }{ \left( P_{ \mathcal{L} }^{T} \left[
u''_T \left(  Y  \right) \right] \right)^2 },
\]
where $X:= \widehat{c}_{ T-1}$, and
\[
Y := \psi_{T} ( \eta_T ( c_0 , ... , c_{T-1} ) ) -
\sum_{l=0}^{T-1} \beta^{(T)}_l c_l  - h_T.
\]
Observe that we have explicitly exploited the fact that $ P_{ \mathcal{L} }^{T}  \left[ \cdot \right] = E \left[ \cdot \ \big| \mathcal{H}_{T-1} \right] $
for markets of type $\mathcal{C}$, and $ P_{ \mathcal{L} }^{T}  \left[ \cdot \right] = E \left[ \cdot \ \big| \sigma \left( \mathcal{G}_{T-1} , \mathcal{F}_{T}  \right) \right] $ for idiosyncratically incomplete markets, by taking the term $ \left(   \frac{ \partial \eta_T }{ \partial c_{T-1} } - \beta^{(T)}_{T-1}  \right)$ out of the brackets.
Furthermore, recall that $ \frac{ \partial \psi_{T-1} }{ \partial W_{T-1} } \geq 0 $ by Theorem \ref{main1}, and note that
$M_k \widetilde{M}_k \geq 0,$ $k=1,...,T$, since $ \frac{M_k}{\widetilde{M}_k} \geq 0 $ in the setting of both markets. Consequently,
in order to prove the statement for $k=T-1$, it suffices to check that
\[
u'''_{T-1} \left( X  \right) P_{ \mathcal{L} }^{T} \left[
u''_T \left(  Y  \right) \right] - u''_{T-1} ( X  ) P_{ \mathcal{L} }^{T} \left[
u''_T \left(  Y  \right) \right] \left(   \frac{ \partial \eta_T }{ \partial c_{T-1} } - \beta^{(T)}_{T-1}  \right) \geq 0,
\]
which is, by (\ref{tec1}), equivalent to
\[
\left(  P_{ \mathcal{L} }^{T} \left[ Y^{-1 - \gamma}  \right]  \right)^2 \leq
X^{ - \gamma } P_{ \mathcal{L} }^{T} \left[ Y^{ - 2 - \gamma }  \right]  \frac{ \widetilde{M}_T }{ \widetilde{M}_{T-1} }.
\]
Notice now that Lemma \ref{reformulation} for $k=T$, implies that the above inequality can be rewritten as
\[
\left(  P_{ \mathcal{L} }^{T} \left[ Y^{-1 - \gamma}  \right]  \right)^2 \leq
P_{ \mathcal{L} }^{T} \left[ Y^{ - 2 - \gamma }  \right] P_{ \mathcal{L} }^{T} \left[ Y^{ - \gamma }  \right] ,
\]
which, by the fact that $  P_{ \mathcal{L} }^{T}  \left[ \cdot
\right] $ is a conditional expectation in the setting of both markets, is satisfied by the
Cauchy-Schwartz inequality. Next, let us treat the case:
$k=T-2$. A differentiation of equation (\ref{tec10}) with respect to
$W_{T-2}$ yields
\[
\frac{ \partial^2 \psi_{T-2} }{ \partial^2 W_{T-2} } = - \frac{1}{ \left( 1 + E \left[  \frac{ M_{ T - 1 } }{ M_{ T - 2 } }  \frac{ \partial \eta_{T-1} }{ \partial c_{T-2} }  \big|  \mathcal{G}_{T-2}  \right]   \right)^2 } E \left[ \frac{M_{T-1}}{M_{T-2}}
\frac{ \partial^2 \eta_{T-1} }{ \partial^2 c_{T-2} }  \frac{ \partial \psi_{T-2} }{ \partial W_{T-2} }    \big| \mathcal{G}_{T-2}  \right] ,
\]
and by differentiating equation (\ref{tec32}) with respect to $c_{T-2}$, we obtain that
\[
\frac{ \partial^2 \eta_{T-1} }{ \partial^2 c_{T-2} } =
\frac{ \widetilde{M}_{T-1} }{ \widetilde{M}_{T-2} }
\frac{ 1  }{ \left( P^{T-1}_{\mathcal{L}} [ u''_{T-1} ( \widehat{c}_{T-1} )   \frac{ \partial \psi_{T-1} }{ \partial W_{T-1} } ] \right)^2 } 
\]
\[
\times
\bigg(  u'''_{T-2} ( \widehat{c}_{T-2} )   P^{T-1}_{\mathcal{L}} \left[ u''_{T-1} ( \widehat{c}_{T-1} )   \frac{ \partial \psi_{T-1} }{ \partial W_{T-1} } \right] -
\]
\[
u''_{T-2} ( \widehat{c}_{T-2} )   P^{T-1}_{\mathcal{L}} \left[ u'''_{T-1} ( \widehat{c}_{T-1} )
 \left(  \frac{ \partial \psi_{T-1} }{ \partial c_{T-2}  } + \frac{ \partial \psi_{T-1} }{ \partial W_{T-1}  }
\frac{ \partial \eta_{T-1} }{ \partial c_{T-2}  } - \beta^{( T-1 )}_{T-2}  \right)  \frac{ \partial \psi_{T-1} }{ \partial W_{T-1} } \right]  -
\]
\[
u''_{T-2} ( \widehat{c}_{T-2} )   P^{T-1}_{\mathcal{L}} \left[ u'''_{T-1} ( \widehat{c}_{T-1} )
\left( \frac{ \partial^2 \psi_{T-1} }{ \partial^2 W_{T-1} }  \frac{ \partial \eta_{T-1} }{ \partial c_{T-2} } +
  \frac{ \partial^2 \psi_{T-1} }{ \partial W_{T-1} \partial c_{T-2} }  \right) \right]
\bigg) .
\]
We shall now simplify the above expression. By plugging (\ref{tec9}) into (\ref{tec6}), and by differentiating the latter equation with respect to $W_{T-1}$, we obtain that
\[
 \frac{ \partial^2 \psi_{T-1}  }{ \partial c_{T-2} \partial W_{T-1} } =
\]
\[
- \frac{ \partial^2 \psi_{T-1}  }{ \partial^2 W_{T-1} } \left( \beta^{(T-1)}_{T-2} + E \left[
\left( \beta^{(T-1)}_{T-2} \frac{ \partial \eta_T }{ \partial c_{T-1} } + \frac{ \partial \eta_T }{ \partial c_{T-2} }  \right) \frac{ M_T }{ M_{T-1} }  \big| \mathcal{G}_{T-1} \right]   \right).
\]
Therefore, we get
\begin{equation}
\frac{ \partial^2 \psi_{T-1} }{ \partial^2 W_{T-1} }  \frac{ \partial \eta_{T-1} }{ \partial c_{T-2} } +
  \frac{ \partial^2 \psi_{T-1} }{ \partial W_{T-1} \partial c_{T-2} } =
\label{tec33}
\end{equation}
\[
\frac{ \partial^2 \psi_{T-1} }{ \partial^2 W_{T-1} } \left( \frac{ \partial \eta_{T-1} }{ \partial c_{T-2} } - \beta^{(T-1)}_{T-2} -
  E \left[
\left( \beta^{(T-1)}_{T-2} \frac{ \partial \eta_T }{ \partial c_{T-1} } + \frac{ \partial \eta_T }{ \partial c_{T-2} }  \right) \frac{ M_T }{ M_{T-1} }  \big| \mathcal{G}_{T-1} \right]
  \right) =
\]
\[
\frac{ \partial^2 \psi_{T-1} }{ \partial^2 W_{T-1} }
\left(
\frac{ \widetilde{M}_{T-1} }{ \widetilde{M}_{T-2} } \frac{ u''_{T-2} ( \widehat{c}_{T-2} )  }{ P_{\mathcal{L}}^{T-1}  \left[ u''_{T-1} ( \widehat{c}_{T-1} ) \frac{ \partial \psi_{T-1} }{ \partial W_{T-1} }   \right] }
\right) ,
\]
where the last equality follows by combining relations (\ref{tec6}), (\ref{tec9}) and (\ref{tec32}). Next, by (\ref{tec32}), it follows that
\[
  P^{T-1}_{\mathcal{L}} \left[ u'''_{T-1} ( \widehat{c}_{T-1} )
 \left(  \frac{ \partial \psi_{T-1} }{ \partial c_{T-2}  } + \frac{ \partial \psi_{T-1} }{ \partial W_{T-1}  }
\frac{ \partial \eta_{T-1} }{ \partial c_{T-2}  } - \beta^{( T-1 )}_{T-2}  \right)  \frac{ \partial \psi_{T-1} }{ \partial W_{T-1} } \right] =
\]
\[
  P^{T-1}_{\mathcal{L}} \left[ u'''_{T-1} ( \widehat{c}_{T-1} )
\left( \frac{ \partial \psi_{T-1} }{ \partial W_{T-1}  } \right)^2  \frac{ \widetilde{M}_{T-1} }{ \widetilde{M}_{T-2} } \frac{ u''_{T-2} \left(
\widehat{c}_{T-2}  \right) }{ P^{T-1}_{ \mathcal{L} } \left[ u''_{T-1} \left( \widehat{c}_{T-1} \right)  \frac{ \partial \psi_{T-1} }{ \partial W_{T-1} }   \right]  }  \right] .
\]
Finally, by combining the above observations, it follows that in order to show that $ \frac{ \partial^2 \psi_{T-2} }{ \partial^2 W_{T-2} } \leq 0 $, it suffices to check that

\[
X^{-2 - \gamma } P^{T-1}_{\mathcal{L}} \left[ Y^{-1 - \gamma}  \frac{ \partial \psi_{T-1} }{ \partial W_{T-1}  }  \right] - X^{-2 - \gamma} \frac{ \widetilde{M}_{T-1} }{ \widetilde{M}_{T-2} } P^{T-1}_{\mathcal{L}} \frac{
\left[ Y^{-2-\gamma}  \left( \frac{ \partial \psi_{T-1} }{ \partial W_{T-1}  }   \right)^2  \right]  }{
\left[ Y^{-1-\gamma}  \frac{ \partial \psi_{T-1} }{ \partial W_{T-1}  }     \right] }  \leq 0 ,
\]
where $X:= \widehat{c}_{T-1} $ and $Y:= \widehat{c}_{T-2} $. By Lemma \ref{reformulation} for $k=T-1$, the latter inequality is equivalent to 
\[
X^{-2 - \gamma } P^{T-1}_{\mathcal{L}} \left[ Y^{-1 - \gamma}  \frac{ \partial \psi_{T-1} }{ \partial W_{T-1}  }  \right] - 
\]
\[
X^{-2 - 2 \gamma}
\frac{ P^{T-1}_{\mathcal{L}}  \left[ Y^{ - \gamma } \right]  }{  X^{- \gamma} }
 \frac{ P^{T-1}_{\mathcal{L}}
\left[ Y^{-2-\gamma}  \left( \frac{ \partial \psi_{T-1} }{ \partial W_{T-1}  }   \right)^2  \right]  }{
P^{T-1}_{\mathcal{L}} \left[ Y^{-1-\gamma}  \frac{ \partial \psi_{T-1} }{ \partial W_{T-1}  }     \right] }  \leq 0 ,
\]
or equivalently
\[
\left( P^{T-1}_{\mathcal{L}} \left[ Y^{-1 - \gamma}  \frac{ \partial \psi_{T-1} }{ \partial W_{T-1}  }  \right]  \right)^2 \leq
 P^{T-1}_{\mathcal{L}}  \left[ Y^{ - \gamma } \right]
P^{T-1}_{\mathcal{L}}
\left[ Y^{-2-\gamma}  \left( \frac{ \partial \psi_{T-1} }{ \partial W_{T-1}  }   \right)^2  \right] ,
\]
which is satisfied by the Cauchy-Schwartz inequality, since the projection $P^{T-1}_{\mathcal{L}}$ is a conditional expectation in both settings of
markets. $\qed$



${}$
\newline
\newline
\textbf{Acknowledgments.}
I am grateful to my supervisor Semyon Malamud for very helpful discussions, and for 
detailed comments on the preliminary version of the manuscript. I would also like to thank 
an anonymous referee for useful remarks. Financial support by the Swiss National Science Foundation via the SNF Grant PDFM2-120424/1 is gratefully acknowledged.

{}

\end{document}